\documentclass[usenatbib,preprint]{mn2e}

\usepackage{graphicx}

\usepackage {natbib,aas_macros}
\usepackage {mediabb}
\usepackage {bm}
\usepackage {amsmath,amssymb}

\newcommand{\msun}{~\thinspace M_\odot} 
\newcommand{\gcm}{~{\rm g~cm}^{-3} }

\newcommand{\Jt}{\tilde{J}}

\newcommand{\vel}{\mathbf{v}}

\newcommand{\cms}{~{\rm cm} ~{\rm s}^{-1} } 
 
\newcommand{\ms}{~{\rm m} ~{\rm s}^{-1} }

\newcommand{\mum}{~{\rm \mu} {\rm m} }

\newcommand{\nm}{~{\rm nm}}
\newcommand{\mm}{~{\rm mm}}

\newcommand{\ccm}{~{\rm cm^{-3}}}

\newcommand{\dv}{\Delta \vel}

\newcommand{\tgrowth}{t_{\rm growth}}

\title{Impact of dust size distribution including large dust grains on magnetic resistivity: an analytical approach}

\author[Tsukamoto et al]{
Yusuke Tsukamoto$^{1}$, Satoshi Okuzumi$^{2}$ \\
$^1$Graduate Schools of Science and Engineering, Kagoshima University, Kagoshima, Japan  \\
$^2$Department of Earth and Planetary Sciences, Kyushu University, Fukuoka, Japan \\
}

\begin{document}
\maketitle

\begin{abstract}
  This paper investigates the impact of dust size distribution on magnetic resistivity. In particular, we focus on its impact
  when the maximum dust size significantly increases from sub-micron.
  The first half of the paper describes our calculation method for magnetic resistivity based on the model
  of \citet{1987ApJ...320..803D} and shows that the method reproduces the results of a more realistic chemical reaction network calculations reasonably well.
  Then, we describe the results of the resistivity calculations for dust distributions with large maximum dust grains.
  Our results show that resistivity tends to decrease with dust growth, which
  is particularly true when the dust size power exponent $q$ is $q=2.5$.
  On the other hand, the decrease is less pronounced when the dust size power exponent $q$ is $q=3.5$,
  i.e., when the small dust is also responsible for the dust cross-section.
  Our results suggest that detailed dust coagulation and fragmentation processes play a vital role in the magnetic resistivities in protostar formation.
\end{abstract}

\begin{keywords}
star formation -- circum-stellar disk -- methods: magnetohydrodynamics -- protoplanetary disk
\end{keywords}

\section{Introduction}
\label{intro}

Non-ideal effects (Ohmic dissipation, Hall effect, and ambipolar diffusion) play a crucial role for formation and evolution of protostars and protoplanetary disks.
For example, Ohmic dissipation and ambipolar diffusion enable formation and stable existence of protoplanetary disks without catastrophic magnetic braking
\citep[e.g.,][]{2011PASJ...63..555M,2015ApJ...810L..26T,2015MNRAS.452..278T,2015ApJ...801..117T, 2016A&A...587A..32M,
  2016MNRAS.457.1037W, 2017ApJ...846....7K, 2017ApJ...835L..11T,2018MNRAS.473.4868Z,2021MNRAS.502.4911X}.
The relatively weak magnetic field of the protostar ($\sim 1$ kG) stems from non-ideal MHD effects
in the first core and disk \citep{2007ApJ...670.1198M,2015MNRAS.452..278T,2018A&A...615A...5V}.
This weak magnetic field around the protostar is the key for protostellar jets formation \citep{2008ApJ...676.1088M,2013ApJ...763....6T,2019ApJ...876..149M}.

The degree of impact of the non-ideal effect depends on the magnetic resistivities, which are determined by the amount of charged particles, and hence ionization chemistry.
In the ionization chemistry, dust grains absorb the ions and electrons and affect their abundance.
The adsorption efficiency depends on the total cross section of dust grains.
Furthermore, a large population of charged small dust grains ($\lesssim$10 \nm)
can contribute to the conductivities \citep{2016MNRAS.460.2050Z}.

The dust size distribution in the previous studies is often assumed to be that of the interstellar medium (ISM)  such as MRN size distribution \citep{1977ApJ...217..425M} or sub-micron sized dust grains.
However, the distribution may change through dust coagulation during protostar formation, particularly in the protoplanetary disks.
By assuming that the relative velocity among the dust is determined by turbulence,
the growth timescale $\tgrowth$ of dust grains in the disk is calculated as \citep{2007A&A...466..413O},
\begin{eqnarray}
  t_{\rm growth} &=&  1.6 \times 10^3  \alpha_{10^{-2}}^{-1/2} \rho_{\rm mat,2 \gcm}^{1/2}  a_{\rm d,  1 \mm}^{1/2} f_{0.01} ^{-1}  \\
  & &n_{\rm g,10^{11} \ccm}^{-1/2}c_{\rm s,190 \ms}^{-1/2}  M_{*, 0.1 \msun}^{-1/4} r_{10 {\rm AU}}^ {3/4} {\rm year}. \nonumber 
\end{eqnarray}
where $n_{\rm g}$, $\alpha$, $\rho_{\rm mat}$, $a_{\rm d}$, $f$, $c_{\rm s}$, $M_*$ denotes the gas number density, viscous $\alpha$ value, material density and size of the dust grains,
dust-to-gas mass ratio, sound velocity, and mass of the central protostar, respectively.  $f_{X}$ means $f_{X}=(\frac{f}{X})$.
Here we assume turbulent velocity and timescale of largest eddy to be $\dv_L=\sqrt{\alpha} c_{\rm s}$ and  $t_L=\Omega^{-1}$, respectively.
Thus, the dust growth timescale is about 100 times smaller than the age of Class 0/I young stellar objects (YSOs),
and dust growth may proceed even in the early evolution of circumstellar disks.
Actually, recent 3D simulation by \citet{2021ApJ...920L..35T} shows that dust growth in the disk (and reflux of large dust to the envelope).
The question we address in this paper is how the change of dust size distribution caused by the dust growth affects resistivities.

Care should be taken in applying the usual chemical-reaction-network calculations to calculate the resistivities with large dust grains because the mean grain charge $\langle Z \rangle$ of $\gtrsim 1 \mum$ is typically \citep{1987ApJ...320..803D}
\begin{eqnarray}
  \langle Z\rangle \sim - 20 a_{\rm d,10 \mum} T_{10 \rm K},
\end{eqnarray}
when abundances of ions and electron is much larger than that of dust grains.
Since dust grains with different charges need to be treated as different chemical species in chemical reaction network calculations,
a vast number of charged dust species should be considered, which is computationally demanding.
Therefore, analytical models of ionization chemistry such as \citet{1987ApJ...320..803D, 2009ApJ...698.1122O,2021ApJ...913..148T}
are more suitable for the calculation of resistivities with large dust grains. Thus, we adopt this approach in this paper.

This paper is organized as follows.
In \S \ref{sec_method}, we describe our analytical model which is based on \citet{1987ApJ...320..803D}.
In \S \ref{sec_validation}, we validate the analytical model by comparing it with chemical reaction calculations.
In \S \ref{sec_results}, we investigate the magnetic resistivities with large dust grains.
Finally, the results are summarized and discussed in \S \ref{discussion}.

\section{Equilibrium charge distribution and magnetic resistivity}
\label{sec_method}
\subsection{Equilibrium of ionization recombination reaction}

We  start from equations for chemical equilibrium in the gas phase.
\begin{eqnarray}
  \label{chemical_equilibrium}
  \zeta n_{\rm g} -s_{\rm i} u_{\rm i} n_{\rm i} \overline{( \sigma_{\rm d}\langle \Jt_{\rm i}(I,Z)\rangle)} n_{\rm d}-\beta n_{\rm i} n_{\rm e}&=&0 \nonumber \\
  \zeta n_{\rm g} -s_{\rm e} u_{\rm e} n_{\rm e} \overline{( \sigma_{\rm d}\langle \Jt_{\rm e}(I,Z)\rangle)} n_{\rm d}-\beta n_{\rm i} n_{\rm e}&=&0
\end{eqnarray}
where
$n_{\rm d}$ is the dust number density.
$ u_{\rm i}=n_{\rm i}^{-1}\sum_k n_{\rm i}^{(k)} u_{\rm i}^{(k)}$,
$ \beta=n_{\rm i}^{-1}\sum_k n_{\rm i}^{(k)} \beta^{(k)}$, and
$ \zeta=n_{\rm g}^{-1}\sum_k n_{\rm g}^{(k)} \zeta^{(k)}$,
are average ion velocity, gas-phase recombination rate coefficient, ionization rate. Here $n_{\rm i}=\sum_k n_{\rm i}^{(k)}$ and $n_{\rm g}=\sum_k n_{\rm g}^{(k)}$
are the total number density of ion and neutral respectively.
$\Jt_{\rm i(e)}(I,Z)$ is the effective cross sections normalized by $\sigma_{\rm d}(I)=\pi (a_{\rm d}(I))^2$ between dust grains and ion (electron). $a_{\rm d}$ is the dust radius.
$\overline{(A)}$ and $\langle A \rangle$ denotes the average over dust size ($I$) and charge ($Z$), respectively.

\citet{1987ApJ...320..803D} derive the approximation formula for the effective cross section of charged particles as, 
\begin{eqnarray}
  \label{Jtilde_mean}
  \Jt(\tau, \nu)=\begin{cases}
  \left(1-\frac{\nu}{\tau}\right)\left[1+\left(\frac{2}{\tau-2\nu}\right)^{1/2}\right] ~(\nu < 0)~ \nonumber \\
  1+(\frac{\pi}{2 \tau})^{1/2}~ (\nu = 0) \\
  \left[1+(4\tau+3\nu)^{-1/2}\right]^2\exp(-\frac{\nu}{\tau(1+\nu^{-1/2})}) ~(\nu > 0),
  \end{cases}
\end{eqnarray}
where $\nu=\nu(Z,q_{\rm i(e)})=Ze/q_{\rm i(e)}$ where $q_{\rm i(e)}$ are charge of ion or electron.
$\tau=\tau(I)= a_{\rm d}(I) k_{\rm B} T/e^2$ is the normalized temperature and  $e$ is the elementary charge.
These formula are correct within few \% for $\tau>10^{-3}$ (thus, the particle size of
$a_{\rm d}>1.67 {\rm nm} (T/(10 K))^{-1}$ which is enough for our purpose).

Using $\Jt(\tau, \nu)$, we obtain,
\begin{eqnarray}
  \label{Jtilde_ion}
  \Jt_{\rm i}(I, Z)=\begin{cases}
  \left(1-\frac{Z}{\tau(I)}\right)\left[1+\left(\frac{2}{\tau(I)-2Z}\right)^{1/2}\right] ( Z < 0) \\
  1+(\frac{\pi}{2 \tau(I)})^{1/2} (Z = 0) \\
  \left[1+(4\tau(I)+3Z)^{-1/2}\right]^2\exp(-\frac{Z}{\tau(I)(1+Z^{-1/2})}) (Z > 0).
  \end{cases}
\end{eqnarray}
for singly charged ions and
\begin{eqnarray}
    \label{Jtilde_electron}
  \Jt_{\rm e}(I, Z)=\begin{cases}
  \left(1+\frac{Z}{\tau(I)}\right)\left[1+\left(\frac{2}{\tau(I)+2Z}\right)^{1/2}\right] ~{\rm for} ~ Z > 0 \\
  1+(\frac{\pi}{2 \tau(I)})^{1/2} ~{\rm for} ~ Z = 0 \\
  \left[1+(4\tau(I)-3Z)^{-1/2}\right]^2\exp(\frac{Z}{\tau(I)(1+(-Z)^{-1/2})}) ~{\rm for} ~ Z < 0.
  \end{cases}
\end{eqnarray}
for electrons.

The charge neutrality condition is given as
\begin{eqnarray}
  \label{charge_neutrality}
  n_{\rm i}-n_{\rm e}+\overline{\langle Z \rangle}n_{\rm d}=0.
\end{eqnarray}

The governing equations for dust charging is the detailed balance equation for dust grains which is given as
\begin{align}
  \label{detailed_balance}
  &n_{\rm i}s_{\rm i}u_{\rm i}  n_{\rm d}(I,Z) \sigma_{\rm d}(I) \Jt_{\rm i}(I,Z)= \nonumber \\
  &n_{\rm e}s_{\rm e}u_{\rm e} n_{\rm d}(I,Z+1) \sigma_{\rm d}(I) \Jt_{\rm e}(I,Z+1) \nonumber \\
  &\therefore \epsilon \frac{n_{\rm d}(I,Z) \Jt_{\rm i}(I,Z)}{n_{\rm d}(I,Z+1) \Jt_{\rm e}(I,Z+1)}=1,
\end{align}
where we define 
\begin{eqnarray}
  \epsilon \equiv \frac{n_{\rm i}s_{\rm i}u_{\rm i}}{n_{\rm e}s_{\rm e}u_{\rm e}}.
\end{eqnarray}

The final governing equation is number density conservation for each $I$
\begin{eqnarray}
  \label{number_density_conservation_Z}
  n_{\rm d}(I)=\sum_Z n_{\rm d}(I,Z).
\end{eqnarray}

We assume that $n_{\rm g}$, $\zeta$, $s_{\rm i(e)}$, $a_{\rm d}(I)$, $\beta$, and $n_{\rm d}(I)$ are known.
Our purpose is to obtain $n_{\rm i}$, $n_{\rm e}$, and $n_{\rm d}(I,Z)$.
In the following three subsections,
we describe the procedure for computing these quantities.

\subsection{Low $\tau$ case}
For $\tau\ll1$, the dust charge  concentrates to $Z=-1,0,1$ \citep{1987ApJ...320..803D}.
Therefore, equations (\ref{detailed_balance}) and (\ref{number_density_conservation_Z}) are reduced to be 

\begin{eqnarray}
    \label{low_temp_case_detailed_balance1}
    \epsilon \frac{n_{\rm d}(I,-1) \Jt_{\rm i}(I,-1)}{n_{\rm d}(I,0) \Jt_{\rm e}(I,0)}&=&1, \\
    \label{low_temp_case_detailed_balance2}
  \epsilon \frac{n_{\rm d}(I,0) \Jt_{\rm i}(I,0)}{n_{\rm d}(I,1) \Jt_{\rm e}(I,1)}&=&1,
\end{eqnarray}
and 
\begin{eqnarray}
  \label{low_temp_number_density_conservation_Z}
  n_{\rm d}(I)=n_{\rm d}(I,-1)&+&n_{\rm d}(I,0)+n_{\rm d}(I,1). 
\end{eqnarray}
Furthermore, we prohibit the transition to $|Z|>1$,
\begin{eqnarray}
  \label{low_temp_case_detailed_balance3}
  n_{\rm d}(I,-1) \Jt_{\rm e}(I,-1)=0, \nonumber \\
  n_{\rm d}(I,1) \Jt_{\rm i}(I,1)=0.
\end{eqnarray}

By using $\Jt_{\rm i}(I,Z=-1)=\Jt_{\rm e}(I,Z=1)=\Jt(\tau(I),\nu=-1)$, we obtain
\begin{eqnarray}
  \label{low_temp_case_number_density}
  n_{\rm d}(I,-1) &=&  \frac{1}{\epsilon} \Omega(I) n_{\rm d}(I,0)   \nonumber\\
  n_{\rm d}(I,1)&=& \epsilon  \Omega(I) n_{\rm d}(I,0) \nonumber\\\nonumber\\
  n_{\rm d}(I,0)&=& \frac{n_{\rm d}(I)}{\Xi(I, \epsilon)}, \nonumber\\
\end{eqnarray}
from equations (\ref{low_temp_case_detailed_balance1}), (\ref{low_temp_case_detailed_balance2}) and (\ref{low_temp_number_density_conservation_Z}). Here we have introduced
\begin{eqnarray}
 \Omega(I) = \frac{\Jt(\tau(I),0)}{\Jt(\tau(I),-1)},  \\
\Xi(I, \epsilon) = \left[\frac{\Omega(I)}{\epsilon}+1+\epsilon \Omega(I) \right].
\end{eqnarray}

By knowing $n_{\rm d}(I,Z)$, we can calculate $\langle Z \rangle$, $\langle \Jt_{\rm i}(\tau, Z) \rangle$, and $\langle \Jt_{\rm e}(\tau, Z) \rangle$  for low $\tau$ case as
\begin{eqnarray}
  \label{low_temp_case_Z_J}
  \langle Z \rangle_{Z, {\rm low}}&=&\frac{\Omega(I)}{\Xi(I,\epsilon)}\left[\epsilon- \epsilon^{-1} \right], \\
  \langle \Jt_{\rm i}(\tau(I)) \rangle_{Z, {\rm low}}&=&\frac{1}{\Xi(I,\epsilon)}\left[\epsilon^{-1} \Omega(I) \Jt_{\rm i}(I,-1) \right. \nonumber \\
     &+& \left. \Jt_{\rm i}(I,0)+ \epsilon \Omega(I) \Jt_{\rm i}(I,1) \right]   \nonumber \\
  &=&\frac{\Jt(I,0) }{\Xi(I,\epsilon)}\left[\epsilon^{-1}+1  \right], \\
  \langle \Jt_{\rm e}(\tau(I)) \rangle_{Z, {\rm low}}&=&\frac{1}{\Xi(I,\epsilon)}\left[\epsilon^{-1} \Omega(I) \Jt_{\rm e}(I,-1) \right. \nonumber \\
    &+& \left. \Jt_{\rm e}(I,0)+ \epsilon \Omega(I) \Jt_{\rm e}(I,1) \right]   \nonumber \\
  &=&\frac{\Jt(I,0) }{\Xi(I,\epsilon)}\left[\epsilon +1 \right],
\end{eqnarray}
where $\Jt_{\rm i}(I,1)=0$ and $\Jt_{\rm e}(I,-1)=0$ have been used.

The results above generalize the low $\tau$ case of \citet{1987ApJ...320..803D}.
If we assume
\begin{eqnarray}
  \label{low_temp_case_detailed_balance_draine}
  \epsilon \frac{n_{\rm d}(I,-1) \Jt_{\rm i}(I,-1)}{n_{\rm d}(I,0) \Jt_{\rm e}(I,0)}&=&1, \nonumber\\
  n_{\rm d}(I)&=&n_{\rm d}(I,-1)+n_{\rm d}(I,0), \nonumber\\
  n_{\rm d}(I,-1) \Jt_{\rm e}(I,-1)&=&0,  \nonumber\\
  n_{\rm d}(I,0) \Jt_{\rm i}(I,0)&=&0,
\end{eqnarray}
(i.e., we only consider $Z=0, -1$), instead of equations (\ref{low_temp_case_detailed_balance2}), (\ref{low_temp_number_density_conservation_Z}), and  (\ref{low_temp_case_detailed_balance3}), and $\Jt(\tau, 0)\sim \sqrt{\pi/(2\tau)}$, $\Jt(\tau, -1)\sim 2/\tau$,
we recover equations (4.11) to (4.13) of \citet{1987ApJ...320..803D}.

\subsection{High $\tau$ case}
The equations in the previous subsection hold as long as the dust charge remains at $Z=\pm 1, 0$.
However, as $\tau$ becomes large (i.e., the temperature of the gas increases or the size of the dust increases),
the typical dust charge becomes  $Z \sim -k_{\rm B} T a_{\rm d}/e^2 =-\tau \ll -1 $ and very small.
Thus, the strategy in the previous section of solving the detailed balancing equations in sequence
is not useful for $Z \ll 1 $ (or $\tau \gg1$)
because we have to consider a large number of detailed balance equations.
On the other hand, for large $\tau$, the dust charge distribution can be treated as a continuous distribution, known to become Gaussian distribution.
Then $\langle Z \rangle , \langle \Jt_{\rm i} \rangle , \langle \Jt_{\rm e} \rangle$ can be obtained analytically \citep{1987ApJ...320..803D,2009ApJ...698.1122O}. 

For $\tau \gg 1$, we can approximate $\Jt_{\rm i}(I, Z)$ and $\Jt_{\rm e}(I, Z)$ of equation (\ref{Jtilde_ion}) and (\ref{Jtilde_electron}) assuming $Z<0$,
\begin{eqnarray}
  \label{Jtilde_ion_high}
  \Jt_{\rm i}(I, Z)=  \left(1-\frac{Z}{\tau(I)}\right)
\end{eqnarray}
for singly charged ion and
\begin{eqnarray}
  \label{Jtilde_ion_high}
  \Jt_{\rm e}(I, Z)= \exp \left(\frac{Z}{\tau(I)}\right)
\end{eqnarray}
for electron.

For $\tau \gg 1$, the solution of detailed balance equation (i.e., equation (\ref{detailed_balance}))
is given as \citep{1987ApJ...320..803D,2009ApJ...698.1122O},
\begin{eqnarray}
  \label{Gaussian}
n_{\rm d}(I,Z)=\frac{n_{\rm d}(I)}{\sqrt{2 \pi \langle \Delta Z^2\rangle}} \exp\left[-\frac{(Z-\langle Z \rangle)^2}{2 \langle \Delta Z^2 \rangle}\right],
\end{eqnarray}
where
\begin{eqnarray}
    \label{Gaussian_Z}
    \langle Z \rangle_{Z, {\rm high}}&=&\psi \tau, \\
    \label{Gaussian_dZ}
  \langle \Delta Z^2 \rangle&=&\frac{1-\psi}{2-\psi} \tau. \\
\end{eqnarray}
The dimensionless parameter $\psi$ is the solution of the equation of 
\begin{eqnarray}
  s_{\rm i} n_{\rm i} u_{\rm i}  \Jt_{\rm i}(I, \langle Z \rangle )&=& s_{\rm e} n_{\rm e} u_{\rm e} \Jt_{\rm e}(I,\langle Z \rangle) \nonumber,\\
  \label{psi_eq}
  \therefore \epsilon (1-\psi)&=&\exp \left( \psi \right),
\end{eqnarray}
and hence, $\psi$ is a function of $\epsilon$. Equations (\ref{Gaussian_Z}) to (\ref{psi_eq}) are derived from the detailed balance equation (\ref{detailed_balance}) with the assumption of $n_{\rm d}(I,Z+1)\sim n_{\rm d}(I,Z)+\partial n_{\rm d}(I,Z)/\partial Z $ and $\Jt_{\rm e}(I,\langle Z \rangle +1) \sim\Jt_{\rm e}(I,\langle Z \rangle)$ \citep[for the detail, see][]{2009ApJ...698.1122O}.

Using these equations, we can calculate $\langle Z \rangle$, $\langle \Jt_{\rm i}(\tau, Z) \rangle$, and $\langle \Jt_{\rm e}(\tau, Z) \rangle$  for high temperature case as,
\begin{eqnarray}
  \label{high_temp_case_Z_J}
  \langle \Jt_{\rm i}\rangle_{Z, {\rm high}} &=&(1-\psi(\epsilon)) \\
  \langle \Jt_{\rm e}\rangle_{Z, {\rm high}} &=&\exp[\psi(\epsilon)].
\end{eqnarray}

\subsection{Number density of ions and electrons}
In the previous two sections we have obtained $\langle Z \rangle$, $ \langle \Jt_{\rm i}\rangle$, and $ \langle \Jt_{\rm i}\rangle$ for each $I$
in the high and low $\tau$ limits.

Following the approach of \citet{1987ApJ...320..803D}, we approximate these values for general $\tau(I)$ as,
\begin{eqnarray}
  \label{general_approx1}
  \langle Z \rangle(I, \epsilon)&=&\langle Z \rangle_{{\rm high}}(I, \psi(\epsilon))+\langle Z \rangle_{\rm {\rm low}}(I, \epsilon), \\
  \langle \Jt_{\rm i} \rangle(I, \epsilon) &=&  \langle \Jt_{\rm i}\rangle_{Z, {\rm high}}(\psi(\epsilon))+ \langle \Jt_{\rm i} \rangle_{Z, {\rm low}}(I, \epsilon), \\
  \label{general_approx2}
  \langle \Jt_{\rm e} \rangle(I, \epsilon)&=&  \langle \Jt_{\rm e}\rangle_{Z, {\rm high}}(\psi(\epsilon))+\langle \Jt_{\rm e} \rangle_{Z, {\rm low}} (I, \epsilon).
\end{eqnarray}
Here we explicitly write the variables of these quantities.

By summing these up for $I$, we can calculate $\overline{\langle Z\rangle}$, $ \overline{\langle \Jt_{\rm i}\rangle}$, and $ \overline{\langle \Jt_{\rm i}\rangle}$  as a function of $\epsilon$.

Then $n_{\rm i}$ and $n_{\rm e}$ are obtained from equation (\ref{chemical_equilibrium}) as a function of $\epsilon$,
\begin{align}
  n_{\rm i}&\equiv n_{\rm i}(\epsilon)=\frac{u_{\rm e} s_{\rm e} \overline{\sigma_{\rm d} \langle \Jt_{\rm e}\rangle}n_{\rm d}}{2 \beta} \\
   &\left(\sqrt{1+\frac{4 \beta \zeta n_{\rm g}}{s_{\rm i} u_{\rm i} s_{\rm e} u_{\rm e} \overline{\sigma_{\rm d} \langle \Jt_{\rm i}\rangle}~ \overline{ \sigma_{\rm d}\langle \Jt_{\rm e}\rangle}n_{\rm d}^2}}-1 \right), \nonumber \\
  n_{\rm e}&\equiv n_{\rm e}(\epsilon)=\frac{u_{\rm i} s_{\rm i} \overline{\sigma_{\rm d} \langle \Jt_{\rm i}\rangle}n_{\rm d}}{2 \beta} \\
  &\left(\sqrt{1+\frac{4 \beta \zeta n_{\rm g}}{s_{\rm i} u_{\rm i} s_{\rm e} u_{\rm e} \overline{\sigma_{\rm d} \langle \Jt_{\rm i}\rangle}~ \overline{ \sigma_{\rm d}\langle \Jt_{\rm e}\rangle}n_{\rm d}^2}}-1\right). \nonumber 
\end{align}

The charge neutrality condition becomes,
\begin{eqnarray}
    \label{gx}
n_{\rm i}(\epsilon)-n_{\rm e}(\epsilon)+n_{\rm d}\overline{\langle Z \rangle}(\epsilon)=0.
\end{eqnarray}
Equations (\ref{gx}) is a nonlinear algebraic equation for $\epsilon$, and we solve this equation with Newton-Raphson method.



\subsection{Conductivity and magnetic resistivity}
Using the $n_{\rm i}$, $n_{\rm e}$, and $n_{\rm d}(I, Z)$ obtained in the previous sections,
the conductivity is calculated as follows \citep{2007Ap&SS.311...35W},
\begin{eqnarray}
  \sigma_{\rm O}&=&\sum_{\rm s} \frac{c}{B} n_{\rm s}q_{\rm s}\beta_{\rm s},\\
  \sigma_{\rm H}&=&-\sum_{\rm s}  \frac{c}{B} \frac{n_{\rm s} q_{\rm s}\beta_{\rm s}^2}{1+\beta_{\rm s}^2},\\
\sigma_{\rm P}&=&\sum_{\rm s}  \frac{c}{B} \frac{n_{\rm s} q_{\rm s} \beta_{\rm s}}{1+\beta_{\rm s}^2} .
\end{eqnarray}
where $\sigma_{\rm O, H, P}$ are the Ohmic, Hall, and Pedersen conductivities, respectively, of the charged species.
\begin{eqnarray}
  \beta_{\rm s}=\frac{q_{\rm s} B}{m_{\rm s} c \gamma_{\rm s} m_{\rm g} n_{\rm g}},
\end{eqnarray}
is the product of the cyclotron frequency
and the collision frequency with the neutral gas.
The subscript ${\rm s}$ denotes the charged species.
Here $n_{\rm s}$ and $q_{\rm s}$ are the number density and charge of the species ${\rm s}$.
  $B$ and $c$ are the magnetic field strength and speed of light, respectively.
$\gamma_{\rm s} =\langle \sigma v\rangle_{\rm s}/(m_{\rm s}+m_{\rm g})$
and $\langle \sigma v\rangle_{\rm s}$ is the collisional momentum transfer rate between species ${\rm s}$ and the neutrals.
$m_{\rm g}$ is the mean mass of the gas.
The momentum transfer rate between neutral and charged species was
calculated using the equations described in \citet{2008A&A...484...17P}.

The conductivities of dust grains are separately calculated from low temperature
and high temperature dust size distribution (equations (\ref{low_temp_case_number_density}) and (\ref{Gaussian})) and then summed up.
This treatment is necessary to correctly calculate the Pedersen conductivity to which both the positively and negatively charged dust grains positively contribute.
This method adds the conductivity of the dust in duplicate, but we confirmed that this does not cause an error because the contribution of dust at higher temperature is small (see \S \ref{sec_validation}).

The Ohmic, Hall, and ambipolar  resistivities are  calculated as
\begin{eqnarray}
\eta_{\rm O}&=&\frac{c^2}{4 \pi}\frac{1}{\sigma_{\rm O}},\\
\eta_{\rm H}&=&\frac{c^2}{4 \pi}\frac{\sigma_{\rm H}}{(\sigma_{\rm H}^2+\sigma_{\rm P}^2)},\\
\eta_{\rm A}&=&\frac{c^2}{4 \pi}\frac{\sigma_{\rm P}}{(\sigma_{\rm H}^2+\sigma_{\rm P}^2)}-\eta_{\rm O}.
\end{eqnarray}

\subsection{Chemical reaction network calculation}
We perform chemical reaction network calculations to compare with the analytical model above.
In the chemical reaction network calculations,
we consider ion species 
${\rm H^+,H_2^+,H_3^+,HCO^+,Mg^+}$ 
${\rm  He^+,C^+,O^+,O_2^+,H_3O^+,OH^+,H_2O^+}$ and neutral species 
${\rm H,H_2, He, CO, O_2, Mg, O, C, HCO, H_2O, OH, N, Fe}$.
We also consider neutral and singly charged dust grains, G$^0$, G$^-$, G$^+$.
We consider cosmic-ray ionization,  gas-phase and dust-surface recombination, and ion-neutral reactions.
We also considered the indirect ionization by high-energy photons emitted by direct cosmic-ray ionization
(described as CRPHOT in the UMIST database).
The initial abundance and reaction rates are taken from the UMIST2012 database \citep{2013A&A...550A..36M}.
We neglect grain-grain collisional neutralization
so that the chemical network calculations are consistent with the analytical model.
The chemical reaction network is solved using the CVODE package \citep{hindmarsh2005sundials}.
We calculate the conductivities using the abundances of charged species in the equilibrium state.

\section{Results}
In this section, we compare the analytical calculation with chemical network calculations
and previous studies to justify the analytical calculations in \S 3.1.
In \S 3.2, we investigate the impact of the dust size
distribution with large dust grains on magnetic resistivity using the analytical calculation.
\subsection{Validation of the analytic model}
\label{sec_validation}

\begin{figure}
  \hspace{-20mm}
  \includegraphics[width=55mm,angle=-90]{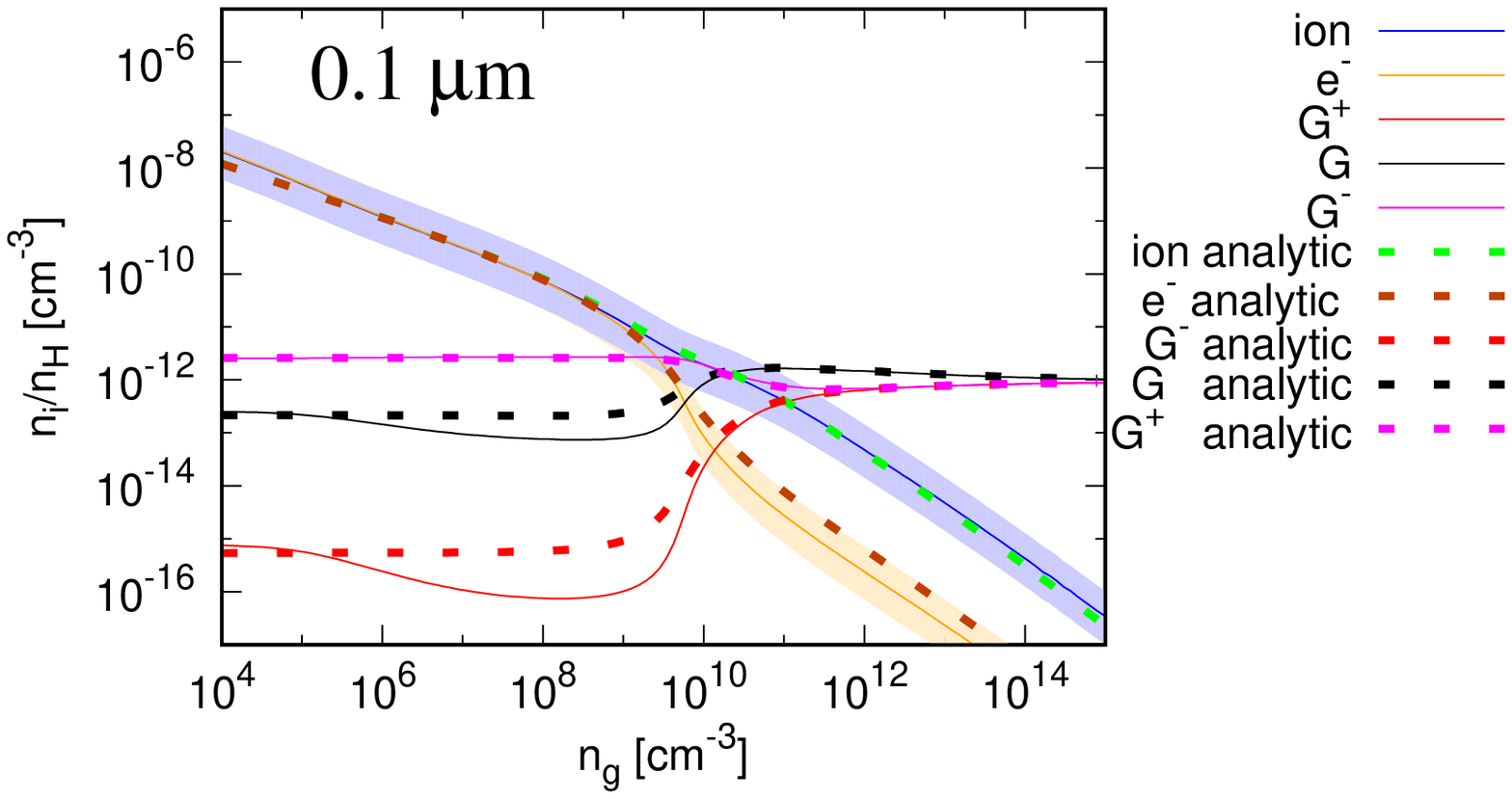} \\
  \hspace{-20mm}
  \includegraphics[width=55mm,angle=-90]{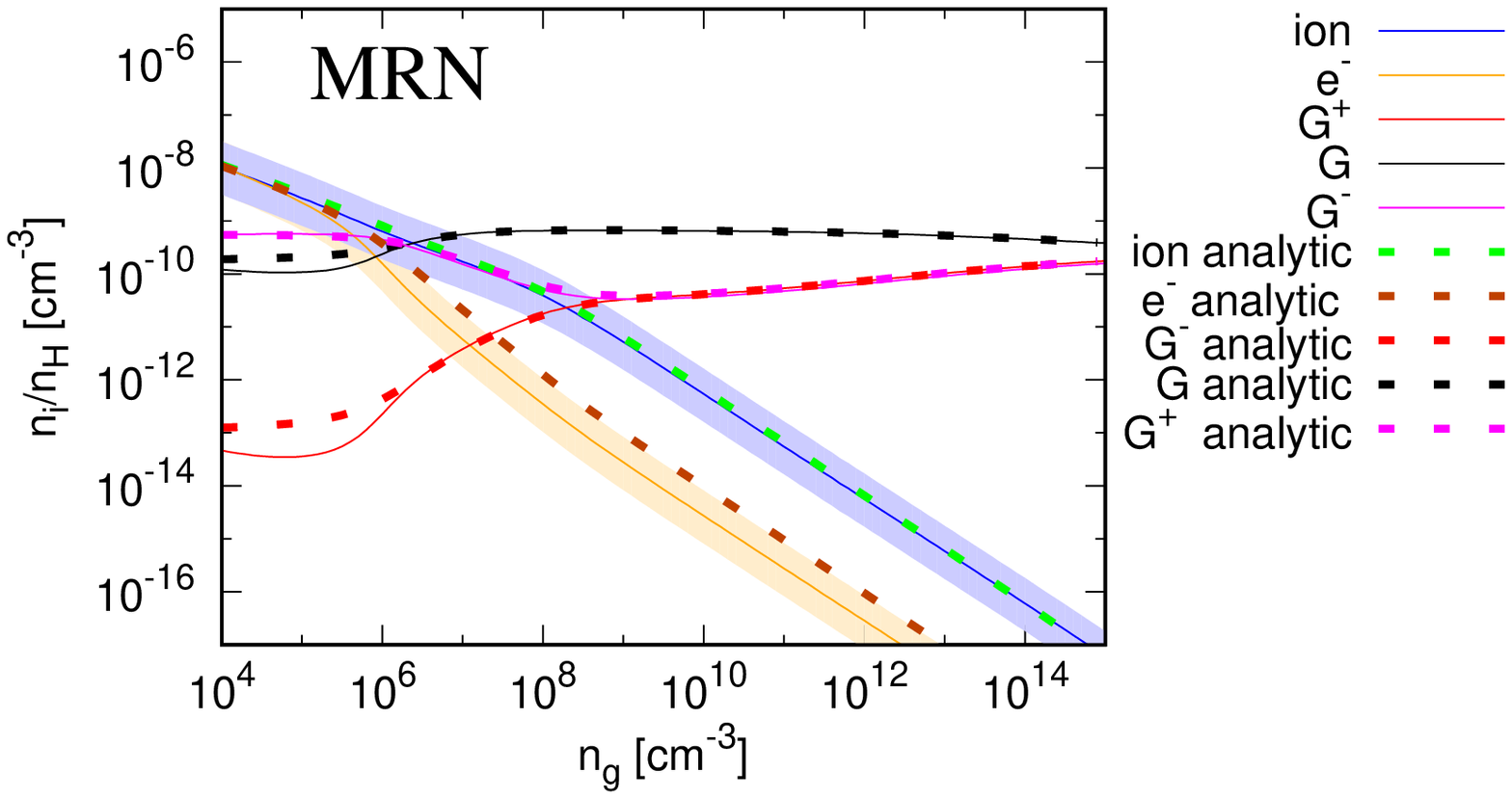}
  \vspace{50mm}
\caption{
  Fractional abundance of ion, electron
  and positively, negatively, and neutral dust grains from the analytic calculation (dotted lines).
  The fractional abundance of total ions, electrons, and charged and neutral dusts from chemical reaction calculations are also plotted with solid lines.
  The orange and blue-shaded regions represent the regions within
  a factor of three of the electron and ion abundances from chemical reaction network, respectively.
  The dust size is assumed to be constant of $a_{\rm d}=0.1 \mum$ in the top panel.
  The dust size distribution is assumed to be MRN size distribution in the bottom panel.
}
\label{abundance}
\end{figure}

\begin{figure*}
  \includegraphics[width=40mm,angle=-90]{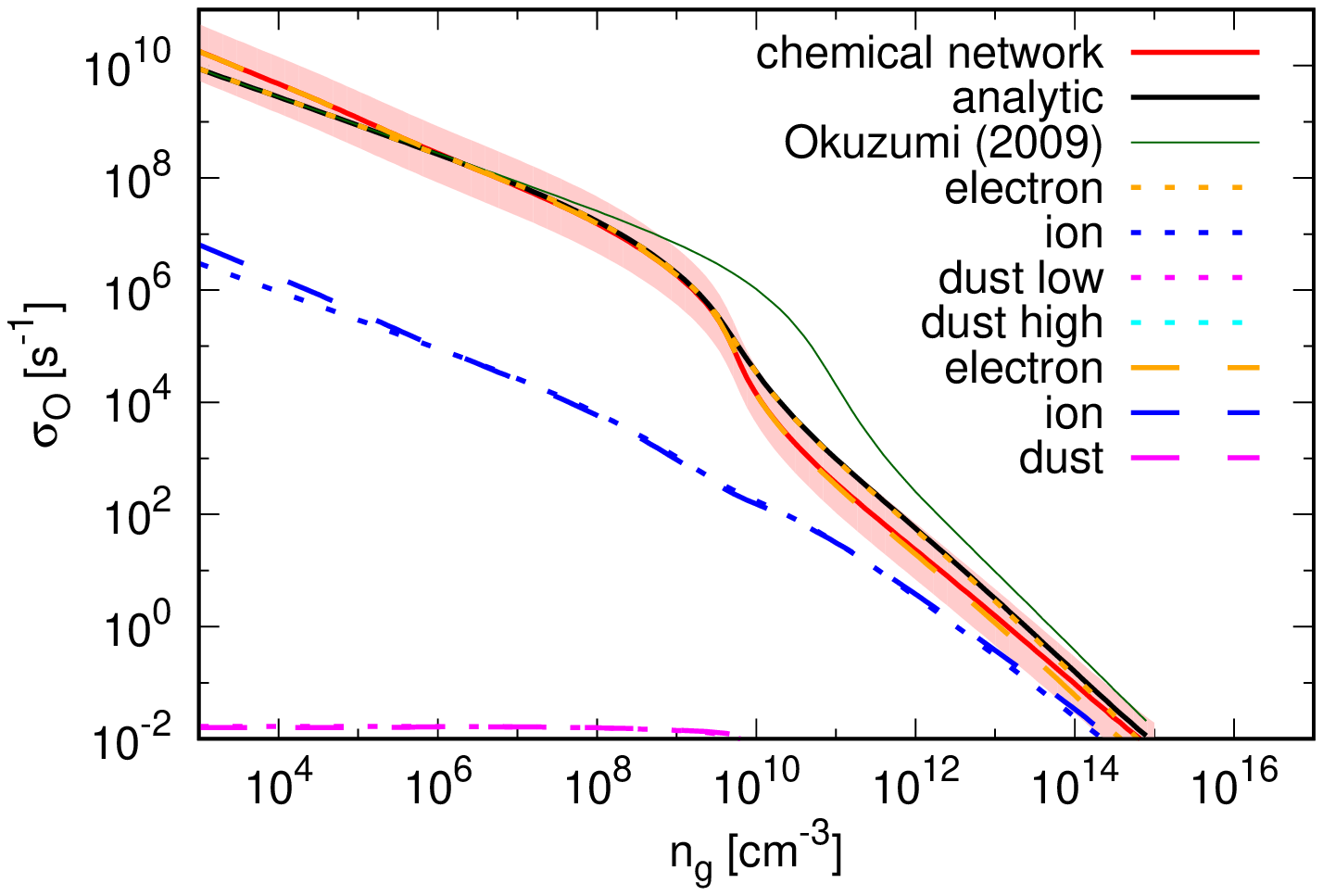}
  \includegraphics[width=40mm,angle=-90]{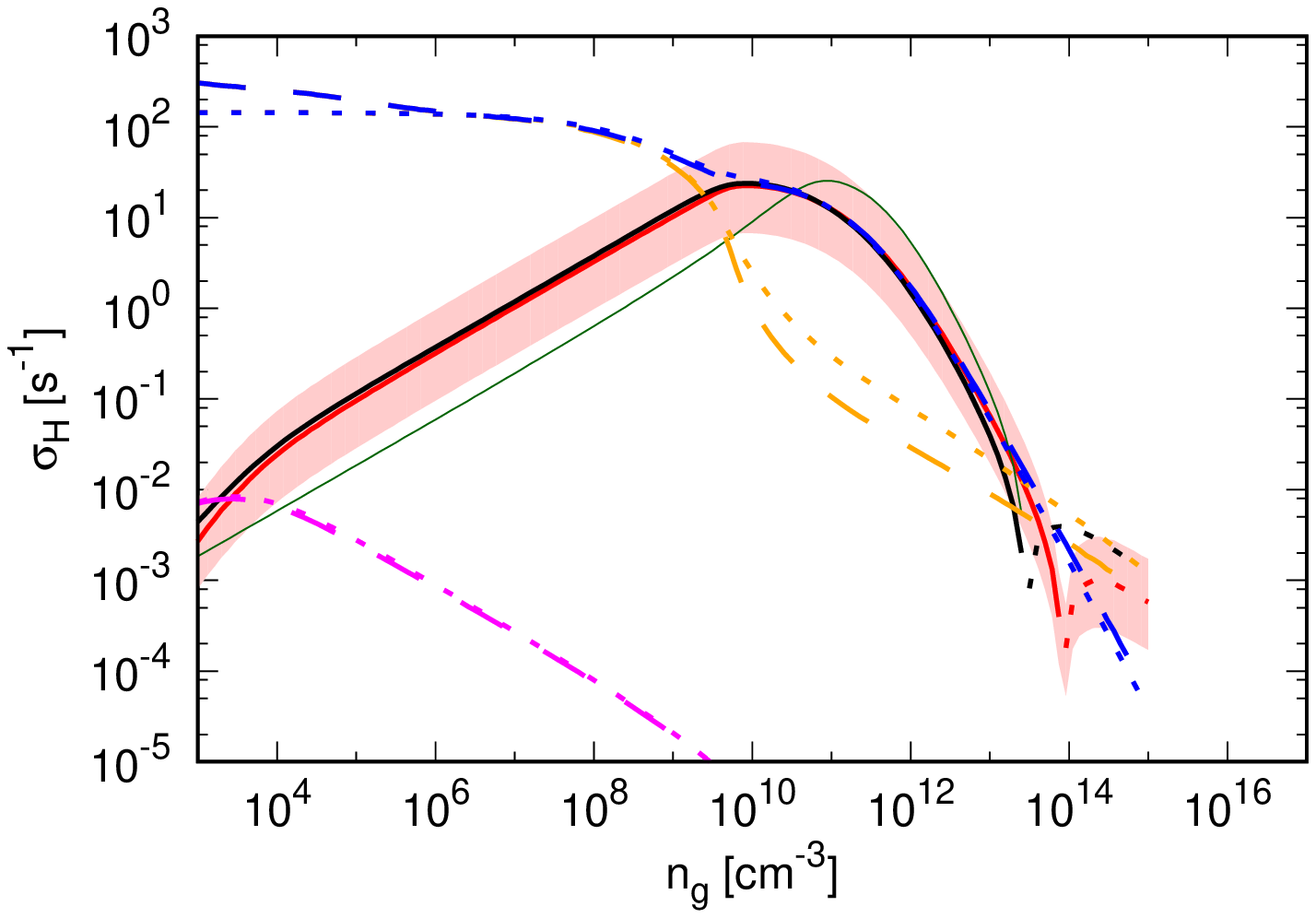}
  \includegraphics[width=40mm,angle=-90]{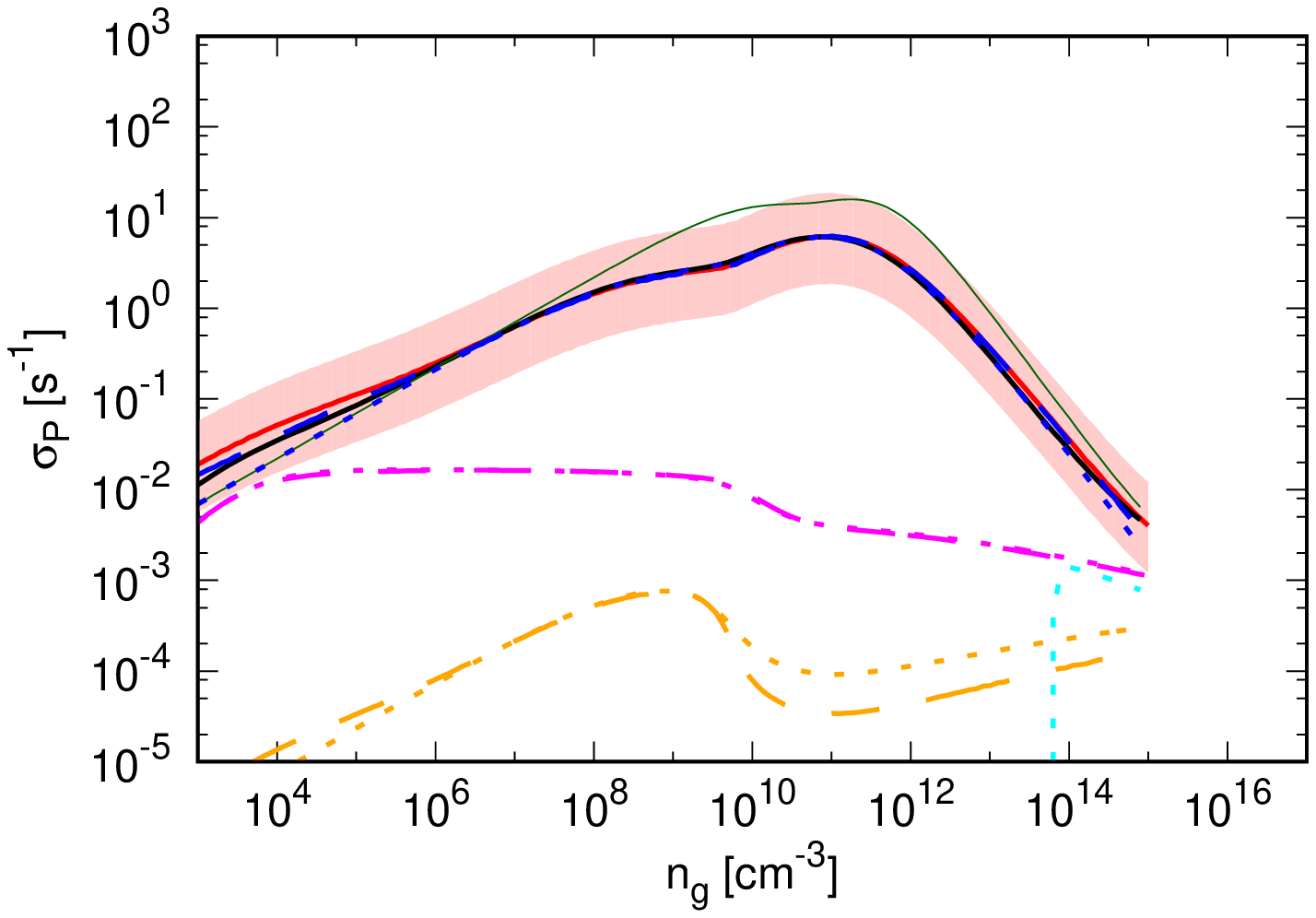}
  \vspace{30mm}
  \caption{
    The Ohmic, Hall, and Pedersen conductivities with $a_{\rm d}=0.1 \mum$ as a function of the density.
    Red and black lines show the total conductivity of the chemical network calculation and the analytic calculation, respectively.
    Dotted red and black lines in the $\sigma_{\rm H}$ plot show that $\sigma_{\rm H}$ is positive.
    The red shaded regions represent the regions within a factor of three of the conductivities of chemical network calculation.
    Dashed and dotted lines other than red and black show the conductivity of the species in chemical network calculation and the analytic calculation, respectively.
    Green lines show the analytic calculations of \citet{2009ApJ...698.1122O}.
}
\label{sigma_01mum}
\end{figure*}

In this subsection, we compare the analytic model with the chemical reaction network calculation.
Here, we assume that the temperature
is $T=10(1+\gamma_T (n_{\rm g}/n_c)^{(\gamma-1)}) ~{\rm K}$, where $\gamma=7/5$ and $n_c=2.6\times 10^{10} \ccm$,
the magnetic field is $0.2 n_{\rm g,\ccm}^{1/2} \mu G$  \citep[i.e., assuming flux freezing; see e.g.,][]{2002ApJ...573..199N},
the dust internal density is $\rho_{\rm mat}=2 \gcm$,
the dust-to-gas mass ratio is $f=0.01$,
and the cosmic ray ionization rate of $\xi_{\rm CR}=10^{-17} {\rm s^{-1}}$ except for the calculations presented in figure \ref{abundance_marchand}.
We assume that the dominant ion is HCO$^+$ and adopt its
recombination rate of $\beta=2.4 \times 10^{-7}(T/300)^{-0.69}$ and its mean molecular weight of $\mu_{\rm I}=29$ for ion in the analytic model.

\subsubsection{Fractional abundances}

Figure \ref{abundance} shows the fractional abundance of ions, electrons and dust grains
for mono-sized dust of $a_{\rm d}=0.1 \mum$ and MRN dust size distribution \citep{1977ApJ...217..425M},
in which the dust size is assumed to be
\begin{eqnarray}
  \label{power_ad}
  \frac{d n_{\rm d}}{d a_{\rm d}} =  A a_{\rm d}^{-q}  (a_{\rm min}<a_{\rm d}<a_{\rm max}),
\end{eqnarray}
where $q=3.5$, $a_{\rm min}=5 \nm$, and $a_{\rm max}=250 \nm$.
$A=(4-q)\rho_{\rm d}/((4/3\pi)\rho_{\rm mat} (\mu_{\rm g}/\mu_{\rm H}) n_{\rm g} |a_{\rm max}^{4-q}-a_{\rm min}^{4-q}|)$ is a constant for normalization.
$\mu_{\rm g}=2.34, \mu_{\rm H}=1.4$, and $\rho_{\rm d}$ is the dust mass density.
$d n_{\rm d}~ d a_{\rm d}$ is the number of dust grains whose sizes are between $a_{\rm d}$ and $a_{\rm d}+da_{\rm d}$ per hydrogen nucleus.

The top panel of figure \ref{abundance} shows the fractional abundance with $a_{\rm d}=0.1 \mum$.
The ion abundance of the analytic model is in good agreement with the chemical reaction network.
On the other hand, our model tends to overestimate the electron abundance in high density region.
The figure shows that ion and electron abundance difference between analytic calculation and chemical
reaction network is within a factor of three.
The negatively charged dust grains G$^-$ and neutral dust grans G$^0$ are dominant in low ($\lesssim 10^{10} \ccm$) and high-density regions ($\gtrsim 10^{10} \ccm$), respectively,
and are in a good agreement between the analytic model and chemical reaction network. Although the abundance of G$^+$ and G$^0$ slightly different between the two calculations around $10^8 \ccm$, this does not cause errors for conductivities.

Bottom panel of figure \ref{abundance} shows the fractional abundance with MRN size distribution.
Even with the size distribution, the similar trend is seen as in the case with $a_{\rm d}=0.1 \mum$, and
the analytic model and chemical reaction network are in a good agreement.

\subsubsection{Conductivity and resistivity}
Figure \ref{sigma_01mum} shows the conductivities from the chemical network calculation and the analytic calculation of this work with $a_{\rm d}=0.1 \mum$.
The figure shows that electrons dominate Ohmic conductivity, so there is about a factor of three discrepancy over the entire region due to differences in the electron abundance.
On the other hand, for the Hall and Pedersen conductivities, the deviation is much smaller than for the Ohmic conductivity because ions dominate them.
As a result, the resulting error for Ohmic resistivity is also about a factor of three, and for Hall and ambipolar resistivity,
the error is smaller than that apart from $\eta_{\rm H}$ of the very low and  high density region as shown in figure \ref{eta_01mum}.
In this figure, we also plot the conductivities obtained by the method of \citet{2009ApJ...698.1122O} which is valid in $\tau \gg 1$.  Since $\tau<1$ almost everywhere in this plot,
we can see that the difference becomes larger when the dust charge affects the ion/electron abundance.

\begin{figure*}
  \includegraphics[width=40mm,angle=-90]{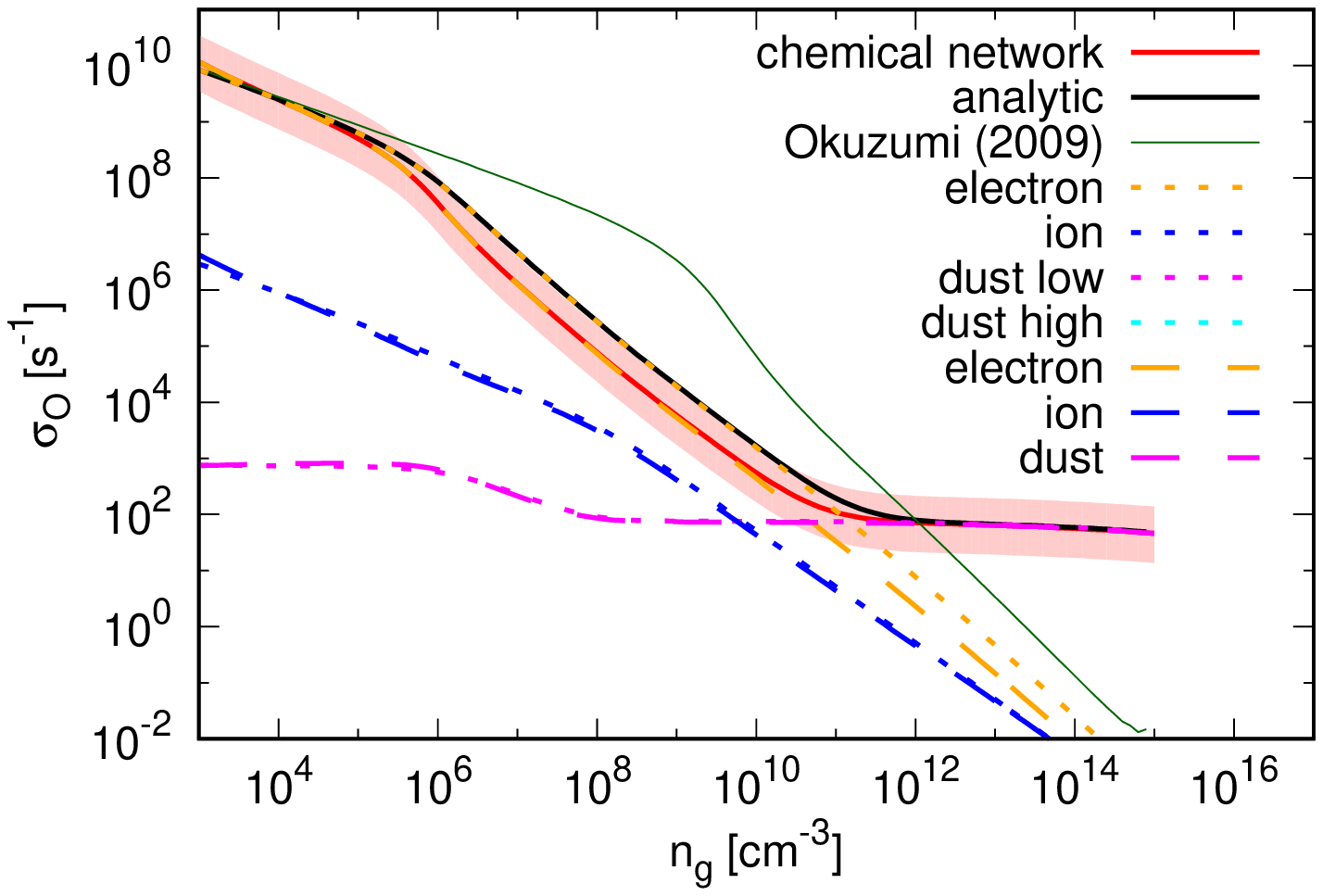}
  \includegraphics[width=40mm,angle=-90]{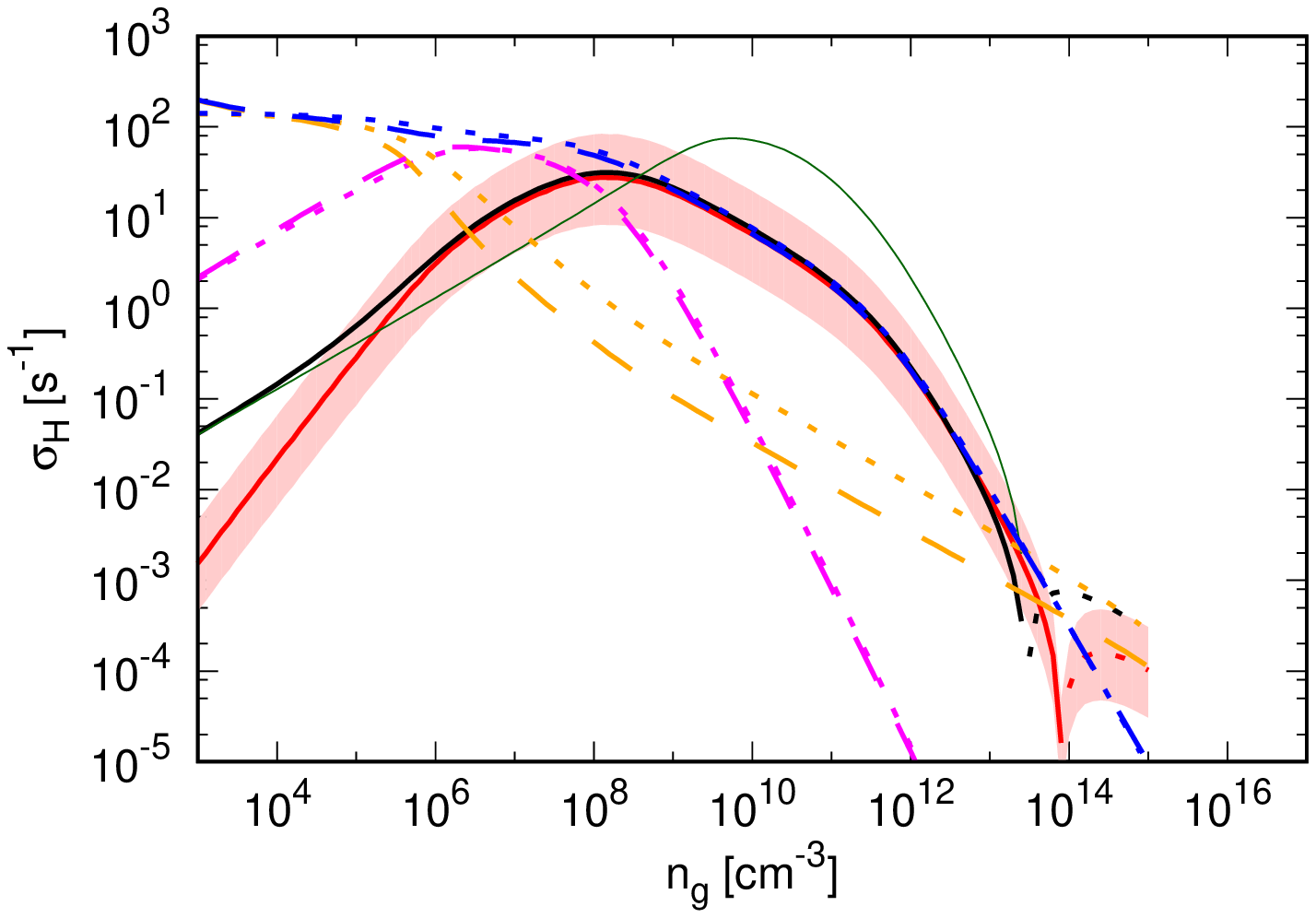}
  \includegraphics[width=40mm,angle=-90]{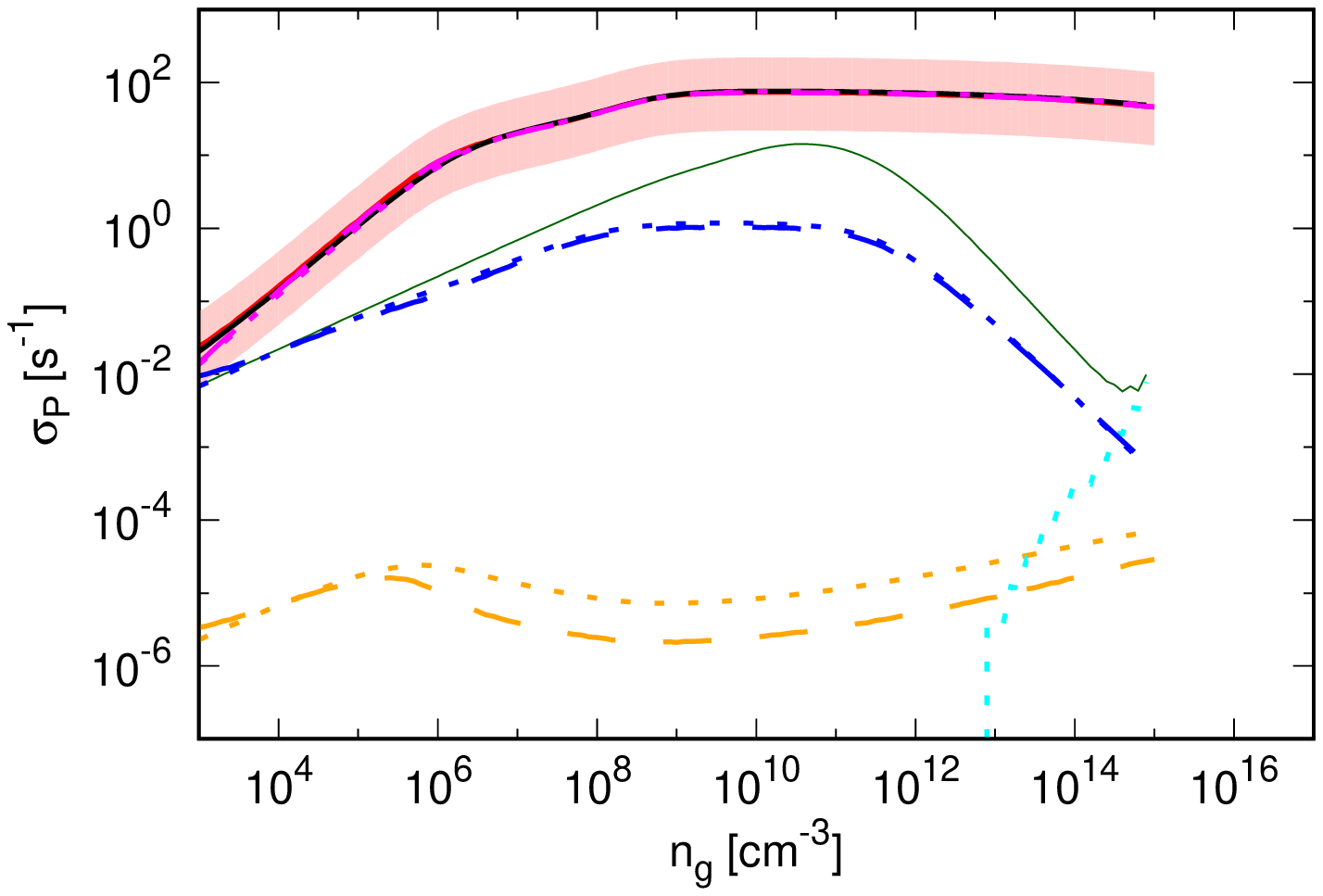}
  \vspace{30mm}
  \caption{
    Same as figure \ref{sigma_01mum} but with MRN dust size distribution.
}
\label{sigma_MRN}
\end{figure*}

\begin{figure*}
  \includegraphics[width=40mm,angle=-90]{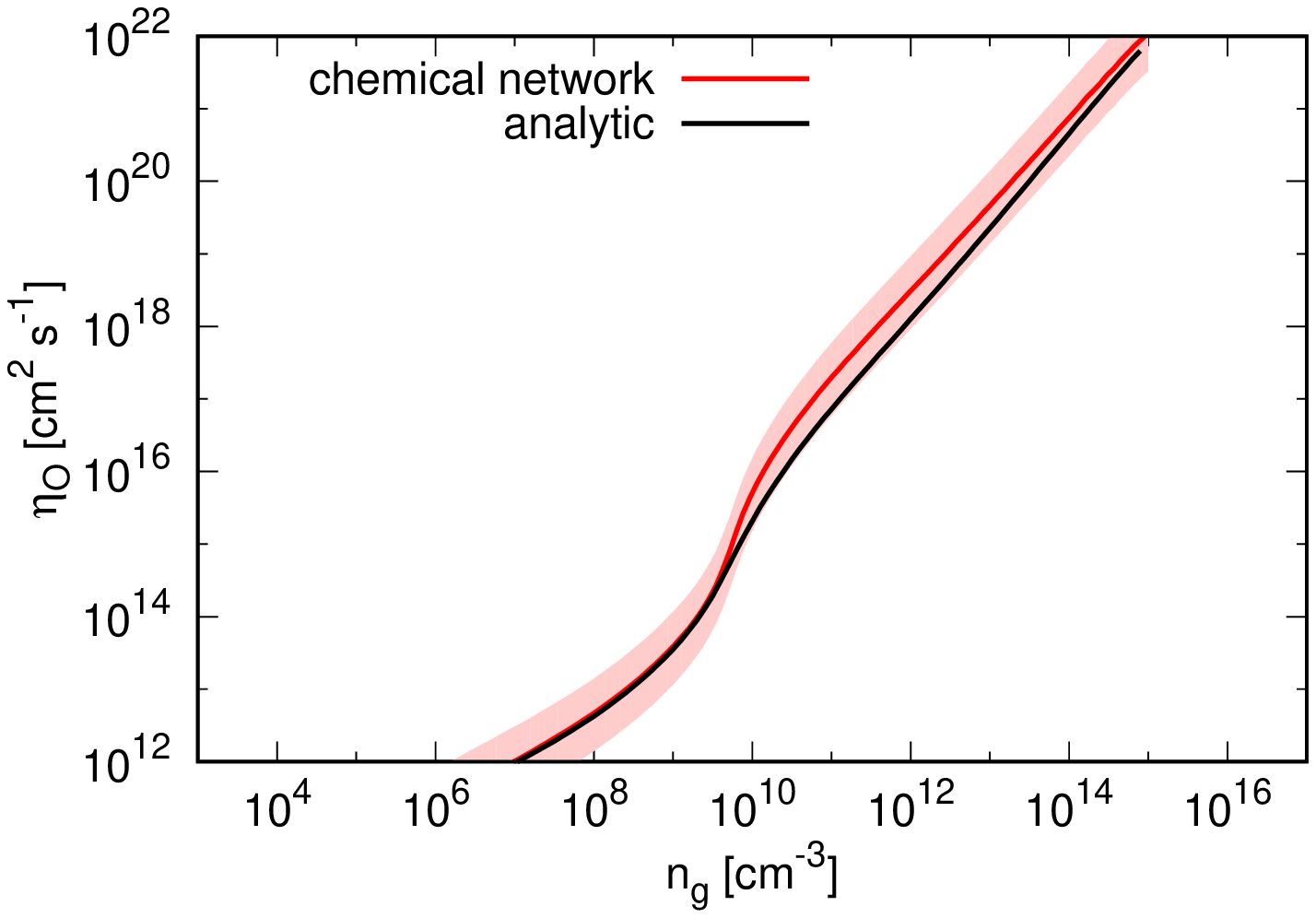}
  \includegraphics[width=40mm,angle=-90]{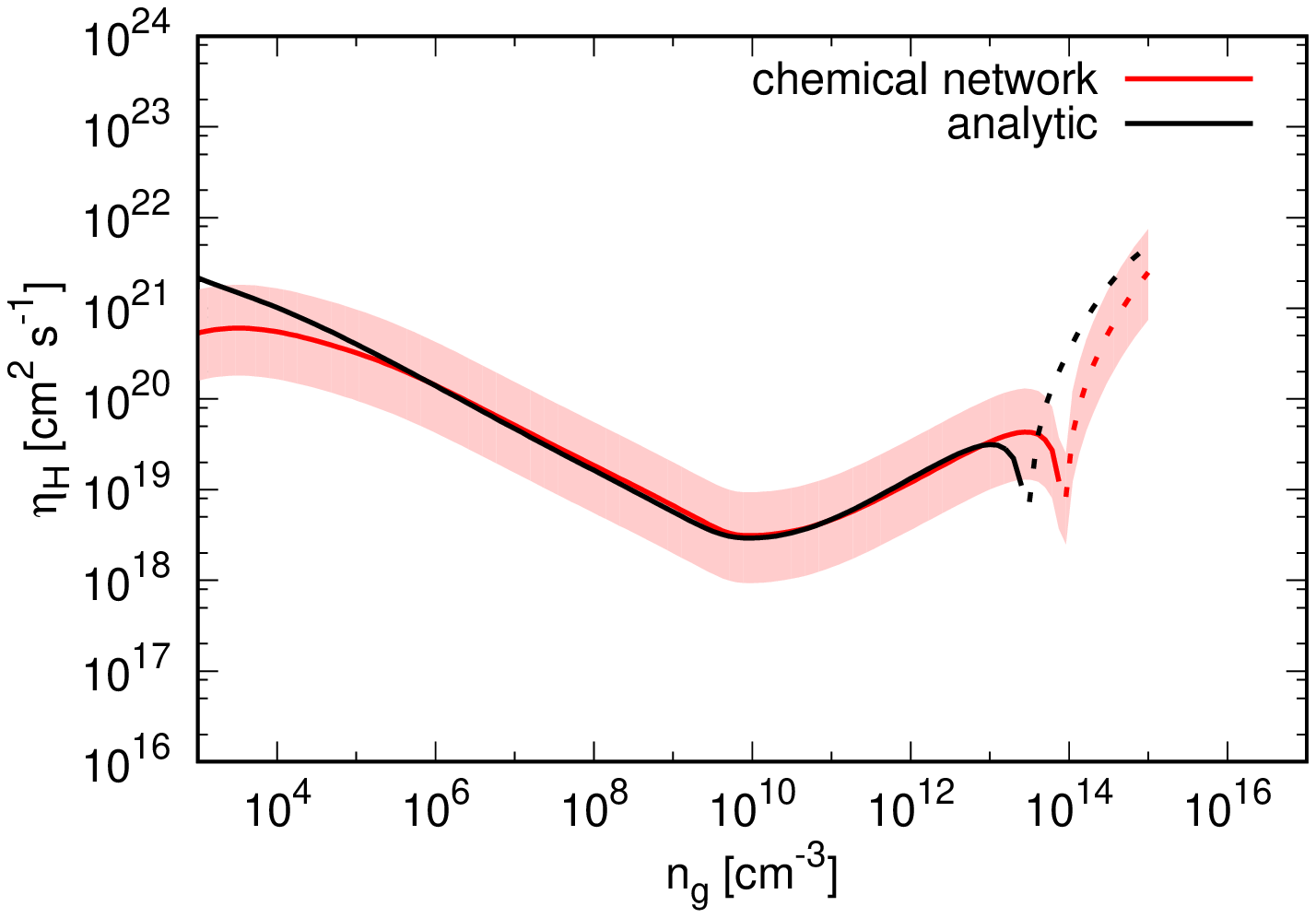}
  \includegraphics[width=40mm,angle=-90]{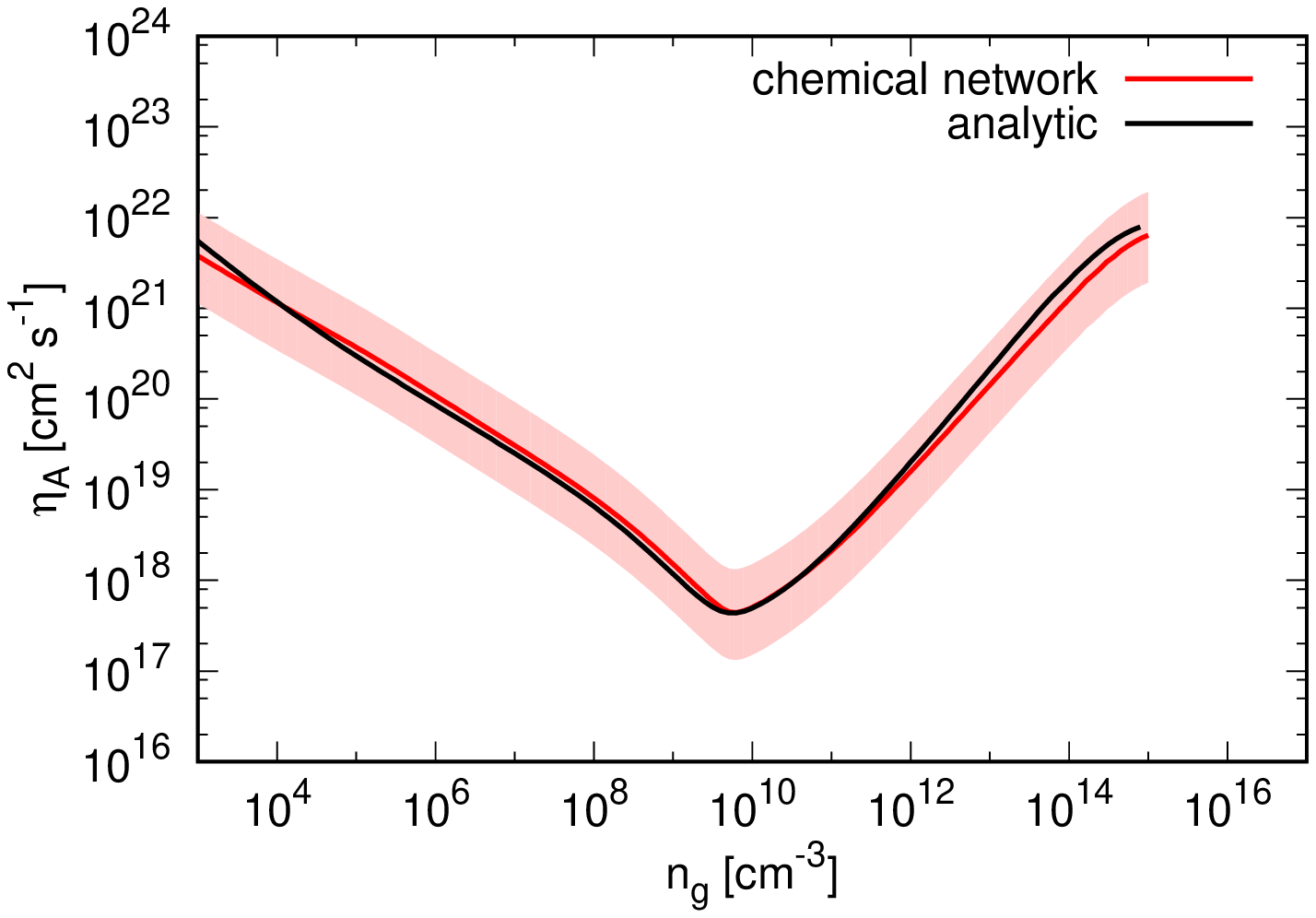}
  \vspace{30mm}
  \caption{
    The Ohmic, Hall, and ambipolar resistivities as a function of the density.
    Dotted red and black lines in the $\eta_{\rm H}$ plot show that $\eta_{\rm H}$ is positive.
    The red shaded regions represent the regions within a factor of three of the resistivities of chemical network calculation.
}
\label{eta_01mum}
\end{figure*}

\begin{figure*}
  \includegraphics[width=40mm,angle=-90]{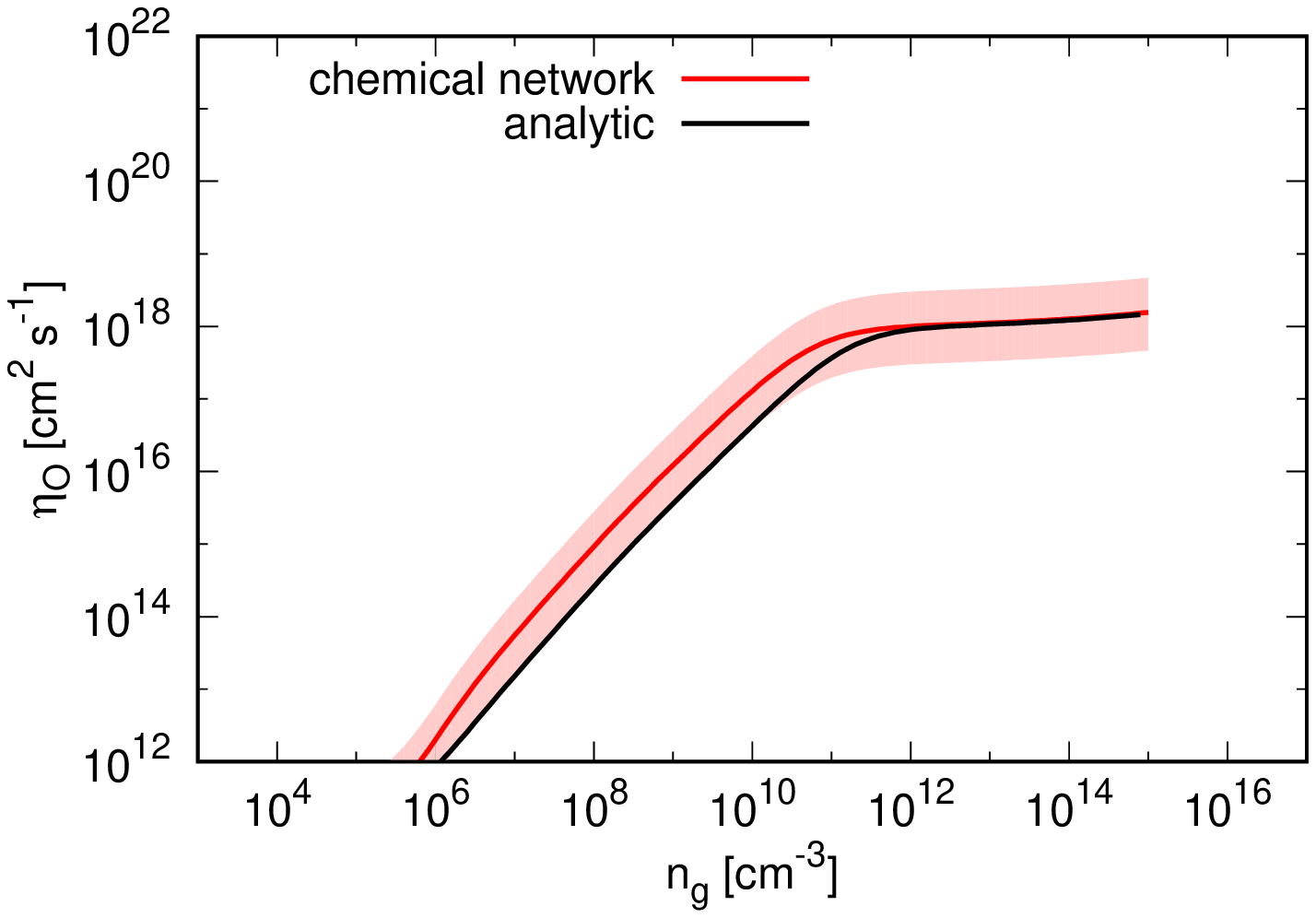}
  \includegraphics[width=40mm,angle=-90]{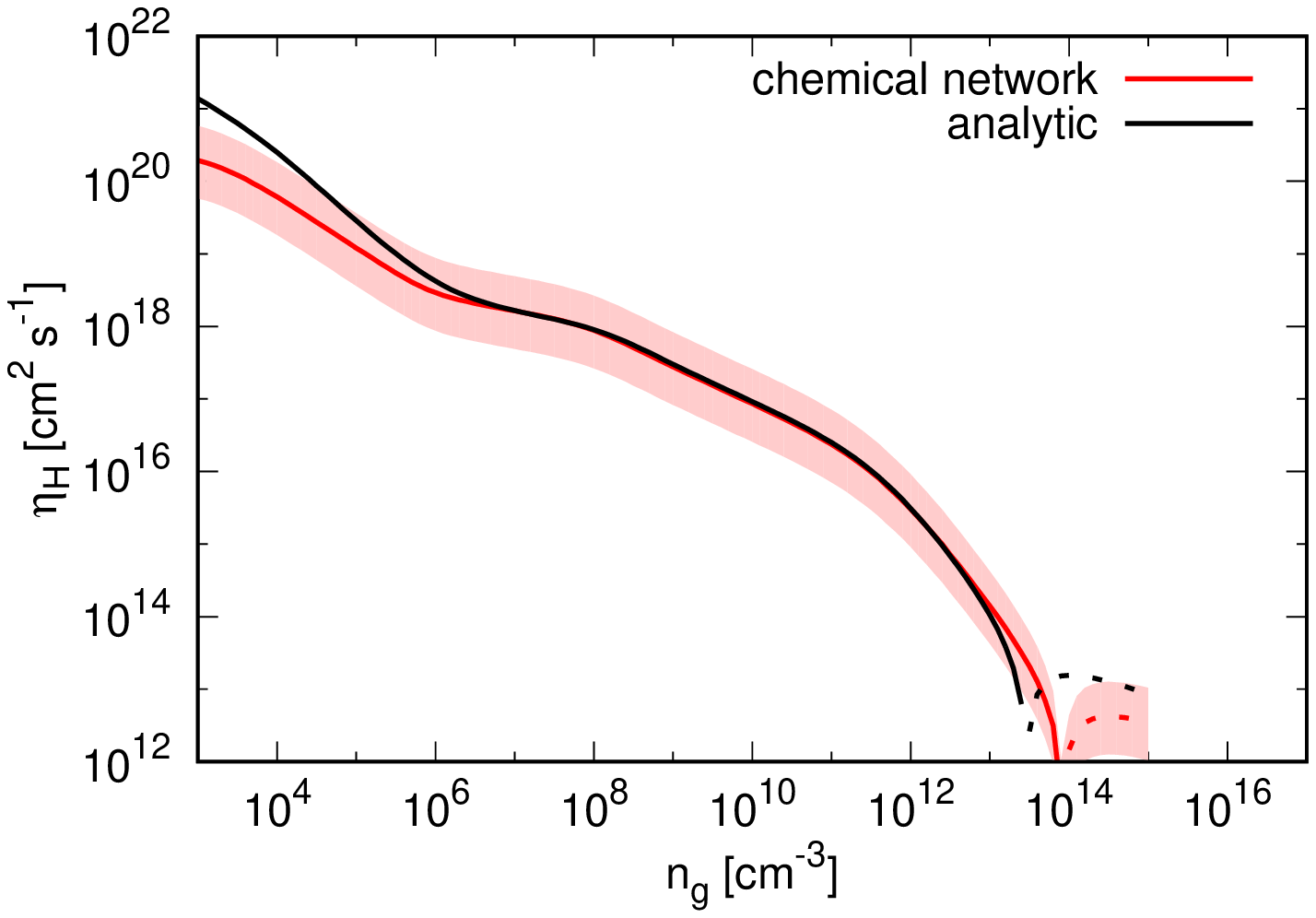}
  \includegraphics[width=40mm,angle=-90]{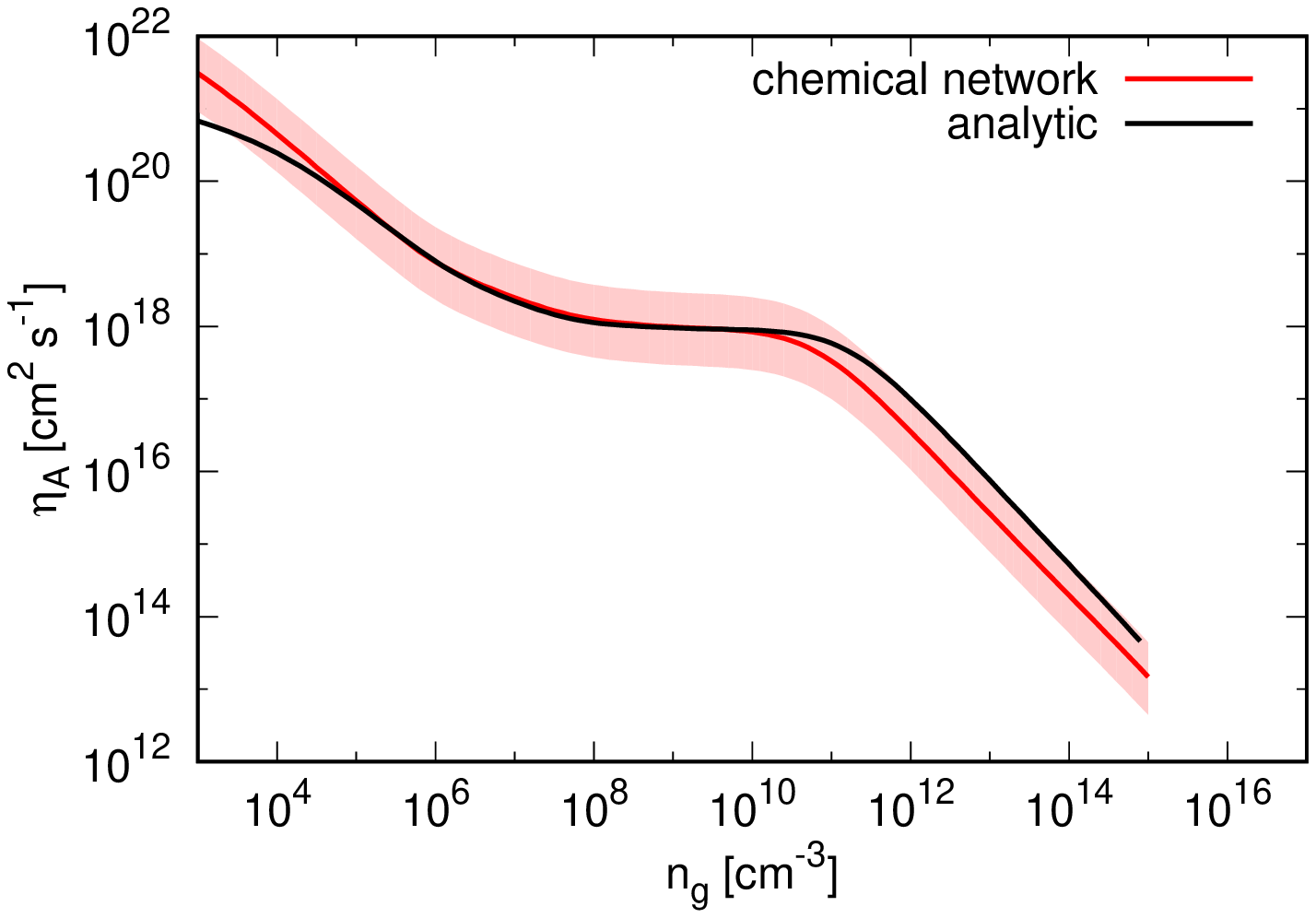}
  \vspace{30mm}
  \caption{
    Same as figure \ref{eta_01mum} but with MRN dust size distribution.
}
\label{eta_MRN}
\end{figure*}

Figure \ref{sigma_MRN} shows the conductivities from the chemical network calculation and the analytic model with MRN size distribution.
In the low-density region of $n_{\rm g} \lesssim 10^{11} \ccm$, Ohmic conductivity is determined by that of electrons as with $a_{\rm d}=0.1 \mum$,
so there is up to about a factor of three discrepancy due to differences in the abundance of electrons.
In the high-density region of $n_{\rm g} \gtrsim10^{11} \ccm$, Ohmic conductivity is determined by that of dust grains, and the discrepancy becomes much smaller.
For Hall conductivity, the deviation is sufficiently small apart from the low-density region of $n_{\rm g}\lesssim10^{5}$ which is due to simplified ion treatment and high-density region of $n_{\rm g}\gtrsim10^{14}$ which is due to the difference of electron abundance.
The deviation is sufficiently small for Pedersen conductivity because they are mainly determined by small dust grains.
As a result, the resulting error for resistivities is also within a factor of three for MRN size distribution apart from the low density region of $n_{\rm g}<10^{-6} \ccm$ where non-ideal MHD effect is not important as shown in figure \ref{eta_MRN}.

\subsubsection{Average dust charge from low to high temperature}
The results in the previous subsection only confirm that the analytical model
is consistent with chemical reaction network calculations
for small $\tau$ (i.e., small dust and low temperature).
In this subsection, we will further check that our model in the high-temperature regime is consistent with \citet{1987ApJ...320..803D}.

\begin{figure}
  \includegraphics[width=40mm,angle=-90]{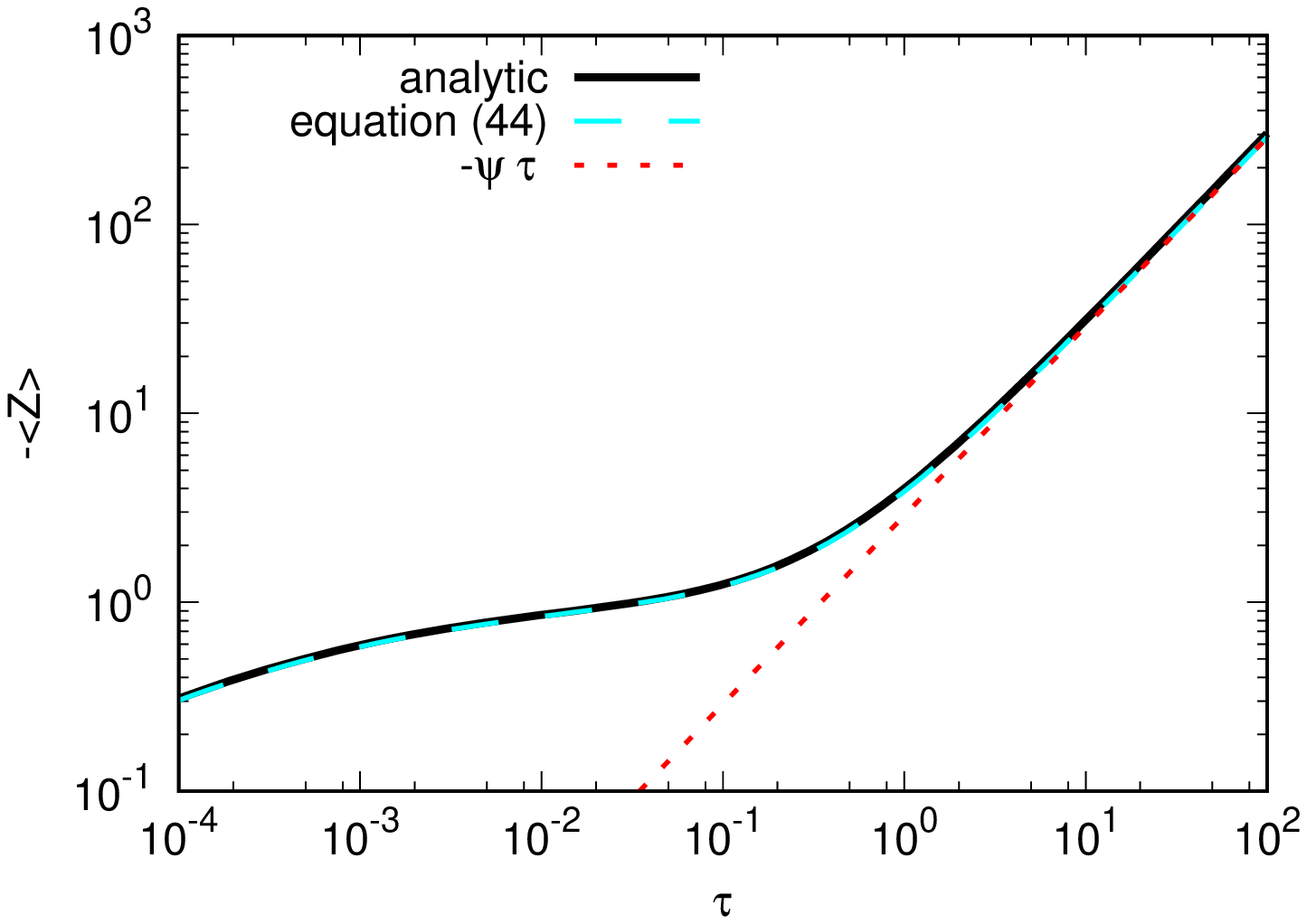}
    \vspace{40mm}
    \caption{
      The mean dust grain charge $-\langle Z \rangle$ as a function of $\tau$.
      In this plot, we set $a_{\rm d}=1 \mum$ and $n_{\rm g}=10^4 \ccm$. 
      The cyan dashed line shows the fitting formula of \citet{1987ApJ...320..803D} (equation (\ref{DS87})).
      The red dotted line shows the $\langle Z \rangle=\psi \tau$, i.e., the mean dust charge of the high-$\tau$ case.
}
\label{Zmean_tau}
\end{figure}

Figure \ref{Zmean_tau} shows the mean dust grain charge $\langle Z \rangle$ as a function of normalized temperature $\tau$.
The cyan dashed line shows the analytic formula of \citet{1987ApJ...320..803D} which is given as
\begin{eqnarray}
  \label{DS87}
\langle Z \rangle=\frac{-1}{1+\sqrt{\tau_0/\tau}}+\psi\tau,
\end{eqnarray}
where $\tau_0=8/(\pi \mu)(m_{\rm e}/m_p)$ and $\mu=(s_{\rm e} n_{\rm e}/n_{\rm i})^2(m_{\rm i}/m_p)$.
The figure shows $\langle Z \rangle$ obeys $\langle Z \rangle\propto \tau$ and our analytic model well reproduces
the analytic formula of \citet{1987ApJ...320..803D}. This means that our model can correctly calculate the ionization
state of dust grains not only for small $\tau$ but also large $\tau$.

\subsection{Impact of dust size on magnetic resistivity}
\label{sec_results}

\begin{figure*}
  \includegraphics[width=40mm,angle=-90]{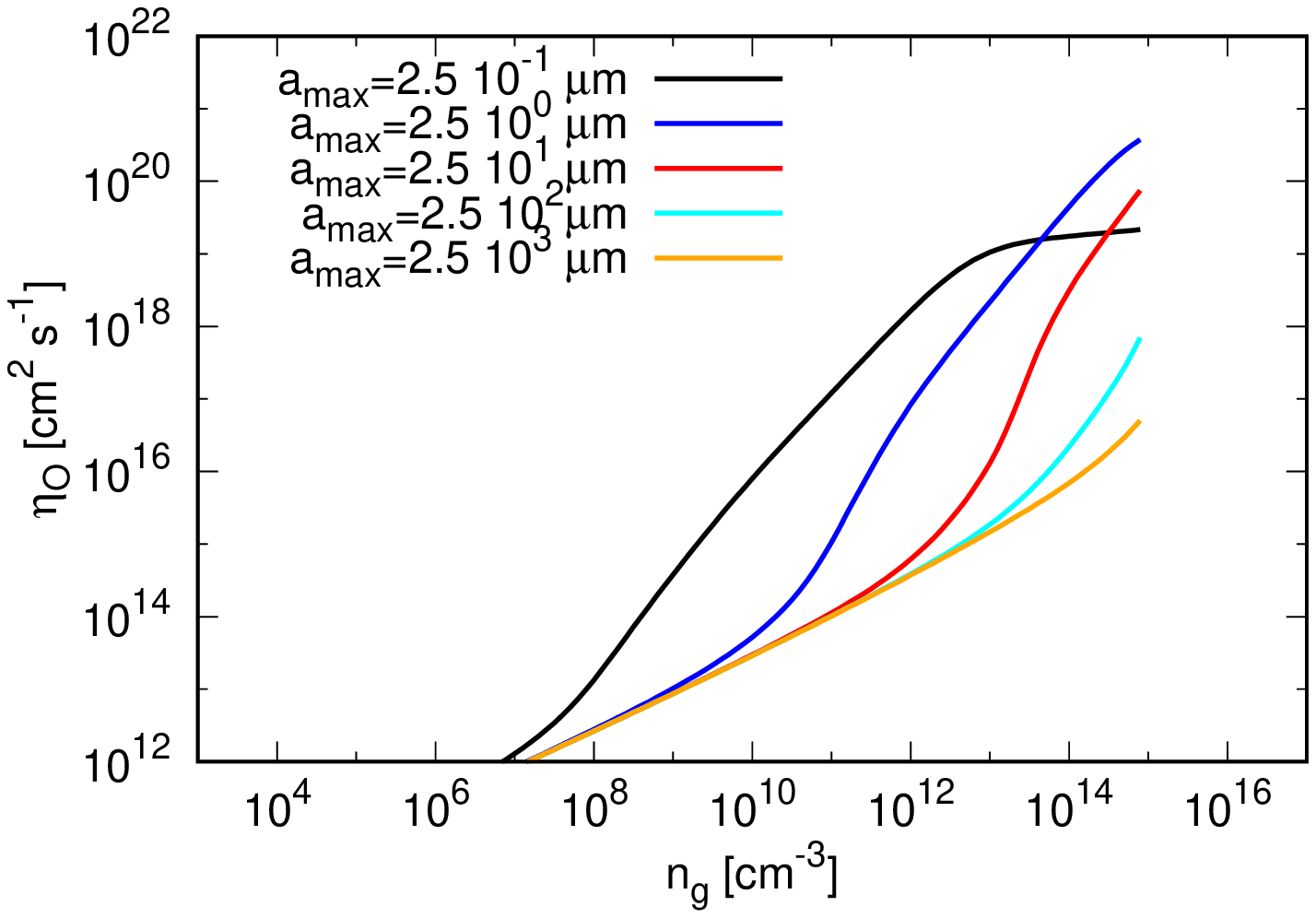}
  \includegraphics[width=40mm,angle=-90]{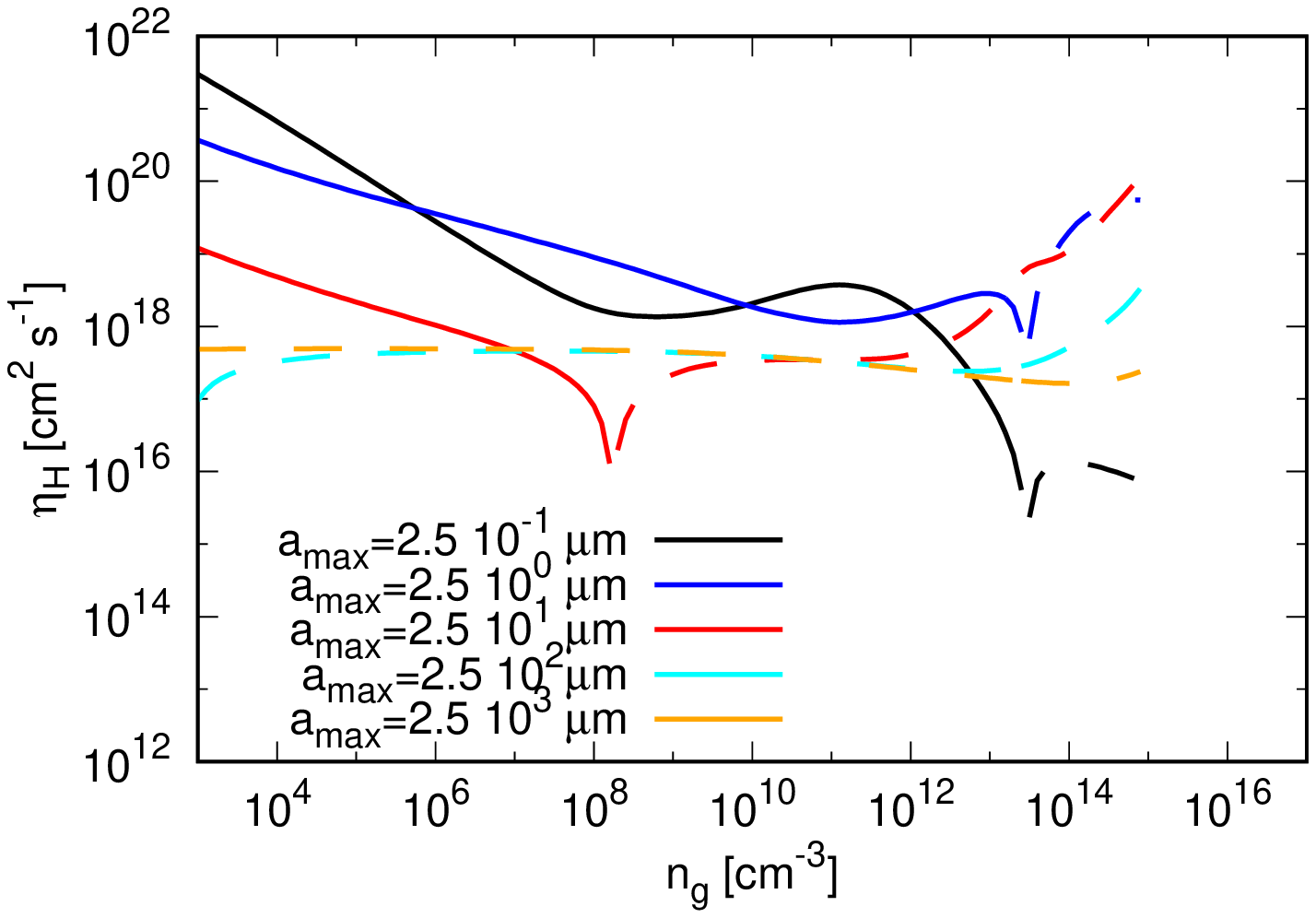}
  \includegraphics[width=40mm,angle=-90]{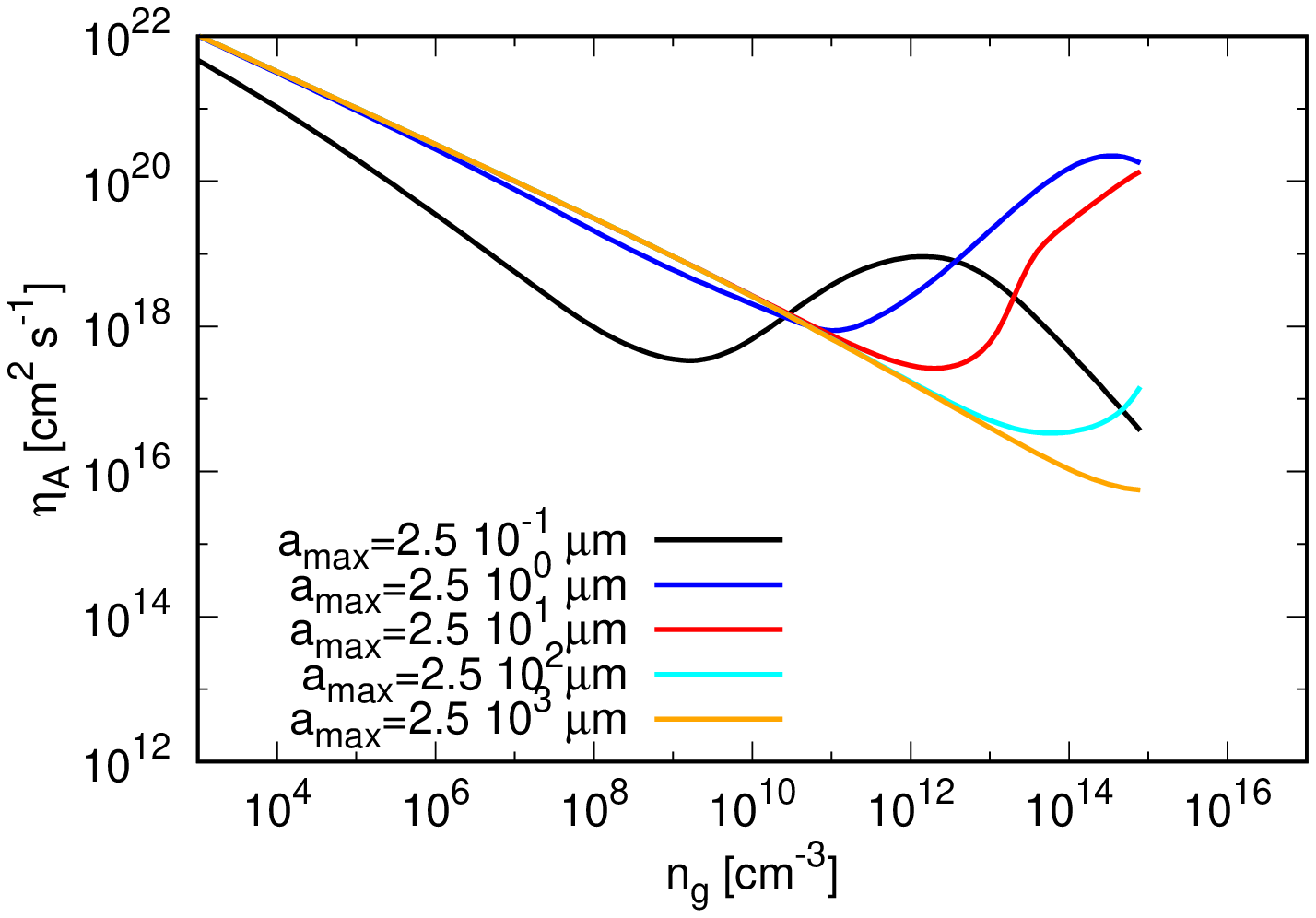}
  \vspace{30mm}
  \caption{
    $\eta_{\rm O}$, $\eta_{\rm H}$, and $\eta_{\rm A}$ with $a_{\rm min}=5 \nm$ and $q=2.5$.
    The black, blue, red, cyan, orange lines show the results of $a_{\rm max}=2.5\times 10^{-1} \mum,2.5\times 10^{0} \mum,2.5\times 10^{1} \mum, 2.5\times 10^{2} \mum, 2.5\times 10^{3} \mum$, respectively.
}
\label{eta_amin5_q25}
\end{figure*}

\begin{figure*}
  \includegraphics[width=40mm,angle=-90]{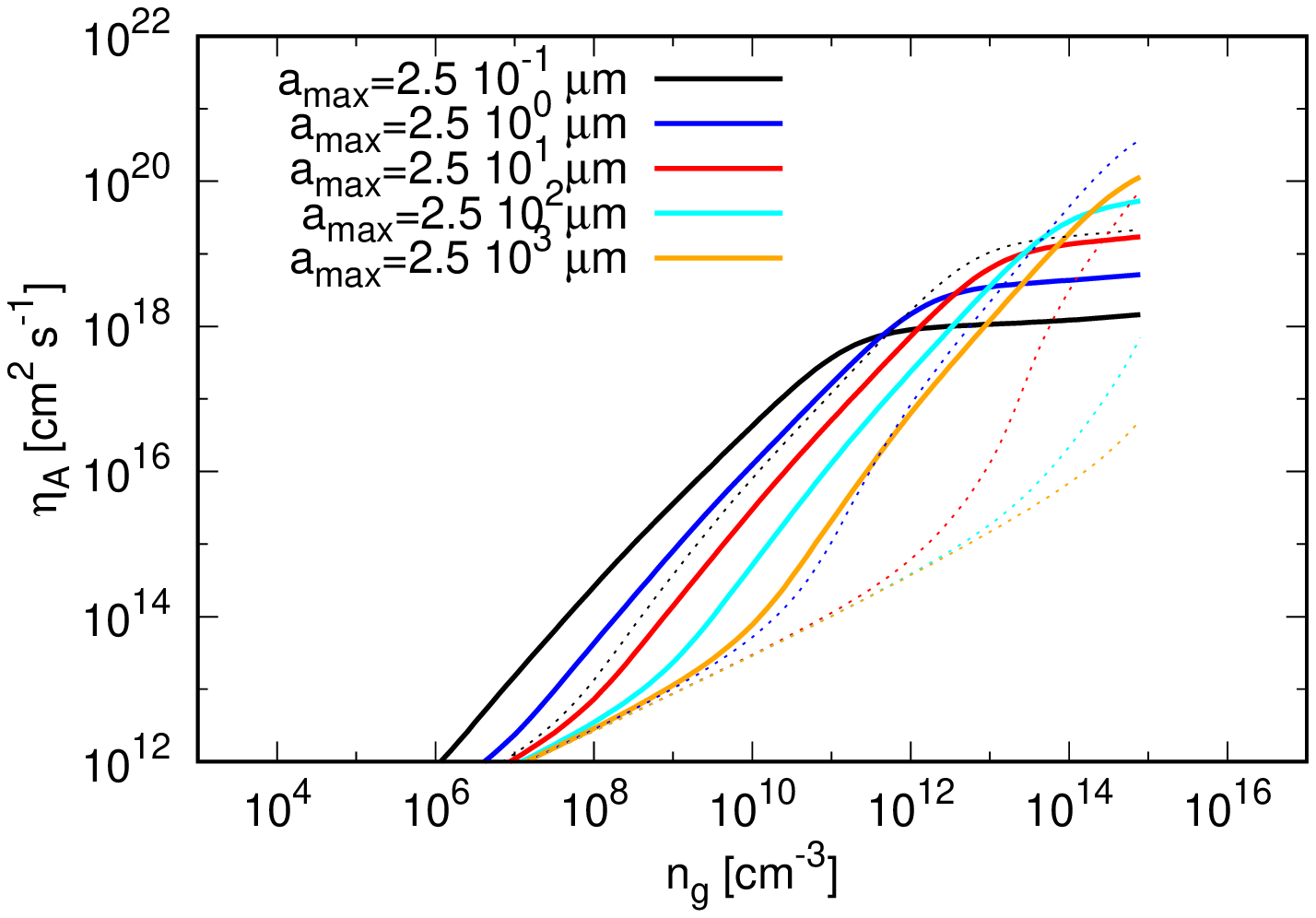}
  \includegraphics[width=40mm,angle=-90]{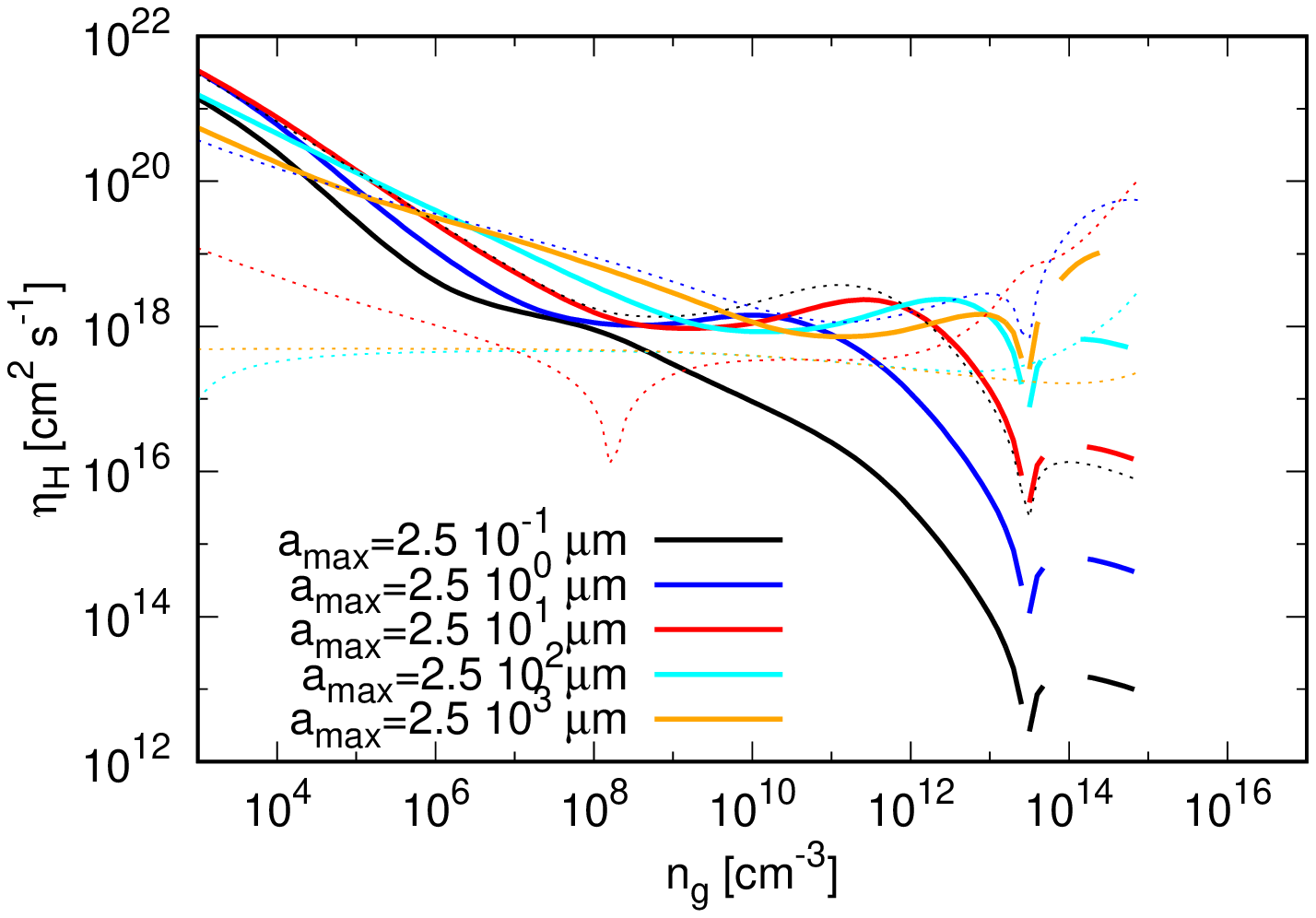}
  \includegraphics[width=40mm,angle=-90]{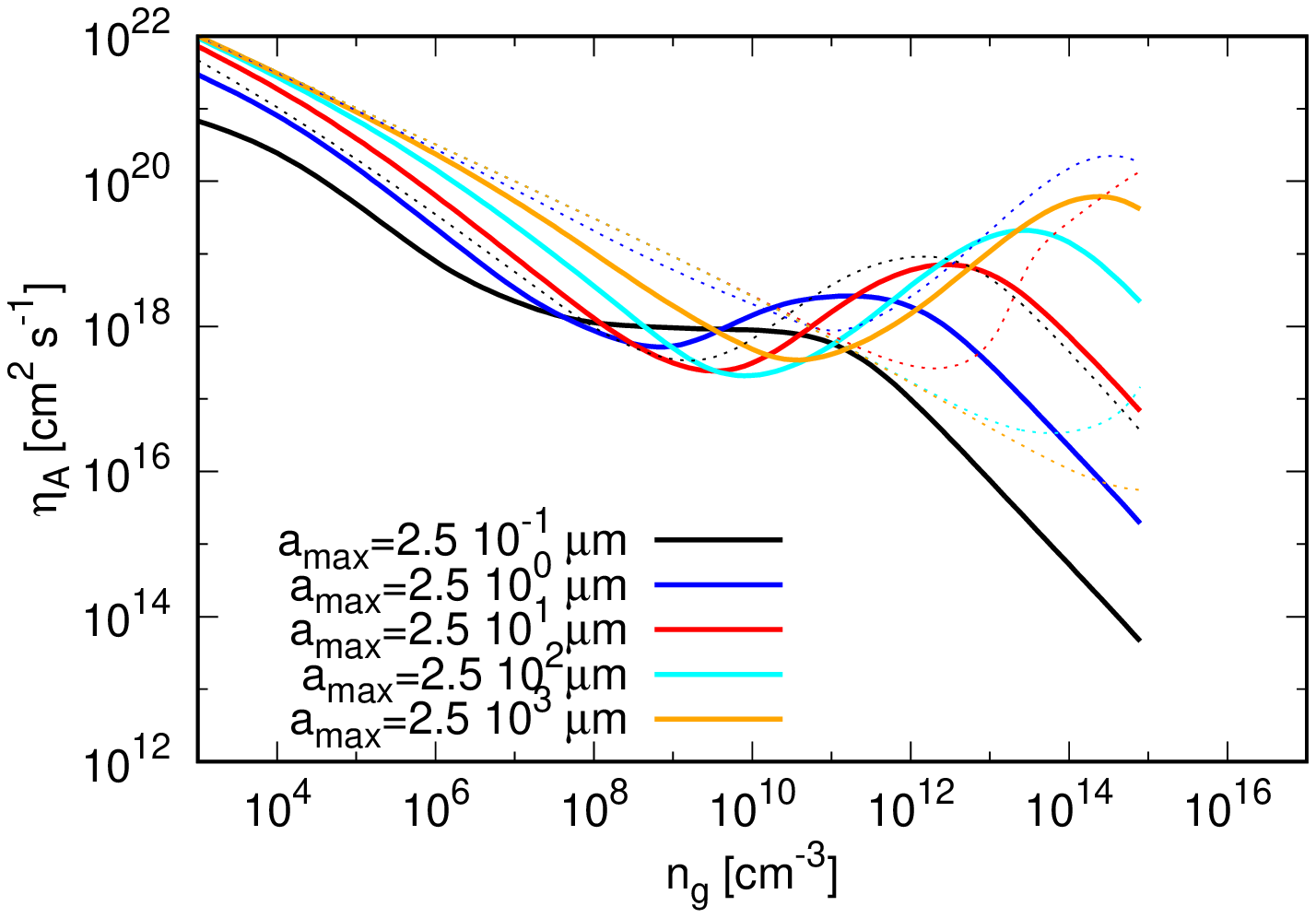}
  \vspace{30mm}
  \caption{
    Same as figure \ref{eta_amin5_q25} but with $a_{\rm min}=5 \nm$ and $q=3.5$.
    Dotted lines show the results with $q=2.5$ for comparison.
}
\label{eta_amin5_q35}
\end{figure*}

In this subsection, we investigate the impact of dust growth on the magnetic resistivities.
We assume power law dust size distribution (equation (\ref{power_ad}))
and vary $a_{\rm min}$, $a_{\rm max}$, and $q$ as parameters.

The important quantity is total cross-section $S_{\rm tot}$ of the dust grains,
\begin{align}
  S_{\rm tot} &= \mu_{\rm H}/\mu_{\rm g} n_{\rm g} A \int_{a_{\rm min}}^{a_{\rm max}} \pi a_{\rm d}^2 a_{\rm d}^{-q} da_{\rm d} \nonumber \\
  &\propto \frac{|{a_{\rm min}}^{-q+3}-{a_{\rm max}}^{-q+3}|}{|{a_{\rm min}}^{-q+4}-{a_{\rm max}}^{-q+4}|},
\end{align}
which determines the absorption efficiency of ions and electrons.
If $4>q>3$ and $a_{\rm max} \gg a_{\rm min}$,   $ S_{\rm tot} \propto {a_{\rm min}}^{-q+3}/{a_{\rm max}}^{-q+4}$ and both small and large dust grains affects the total cross section.
In this case, it is expected that the impact of dust growth would be less significantly on the absorption efficiency of ions and electrons, and hence resistivities.
On the other hand, if $q<3$  and $a_{\rm max} \gg a_{\rm min}$,  $ S_{\rm tot} \propto {a_{\rm max}}^{-1}$ and only large dust grains determines the total surface area.
Thus, it is expected that dust growth would significantly change the resistivities for $q<3$.
It is pointed out that $q\sim 2.5$ when the coagulation process dominates while $q\sim 3.5$ when the disruption process dominates \citep{1993Icar..106...20M}.
Thus, we investigate the resistivities with $q=2.5$ and $q=3.5$.

On the other hand, a large population of charged small grains ($\lesssim$10 \nm) can be responsible for the conductivities \citep[][see also figure \ref{sigma_MRN}]{2016MNRAS.460.2050Z}, and decreases the resistivities at envelope and disk \citep{2018MNRAS.478.2723Z,2019MNRAS.484.2119K,2020ApJ...900..180M, 2020ApJ...896..158T}. 
Thus, whether such small grains exist or not would also be important.
In this subsection, we investigate the resistivities with $a_{\rm min}=5 \nm$ and $a_{\rm min}=0.1 \mum$.

\begin{figure*}
  \includegraphics[width=40mm,angle=-90]{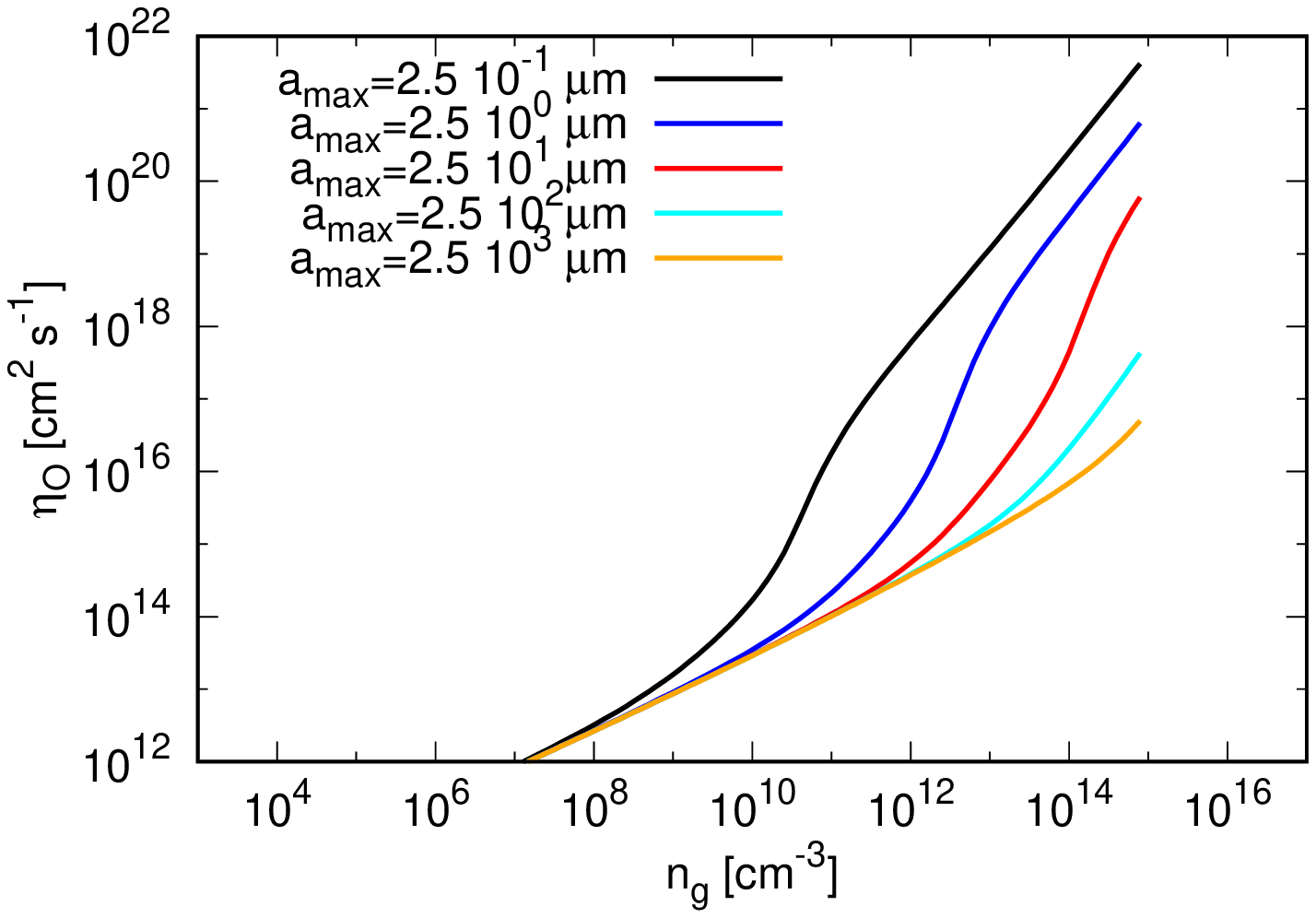}
  \includegraphics[width=40mm,angle=-90]{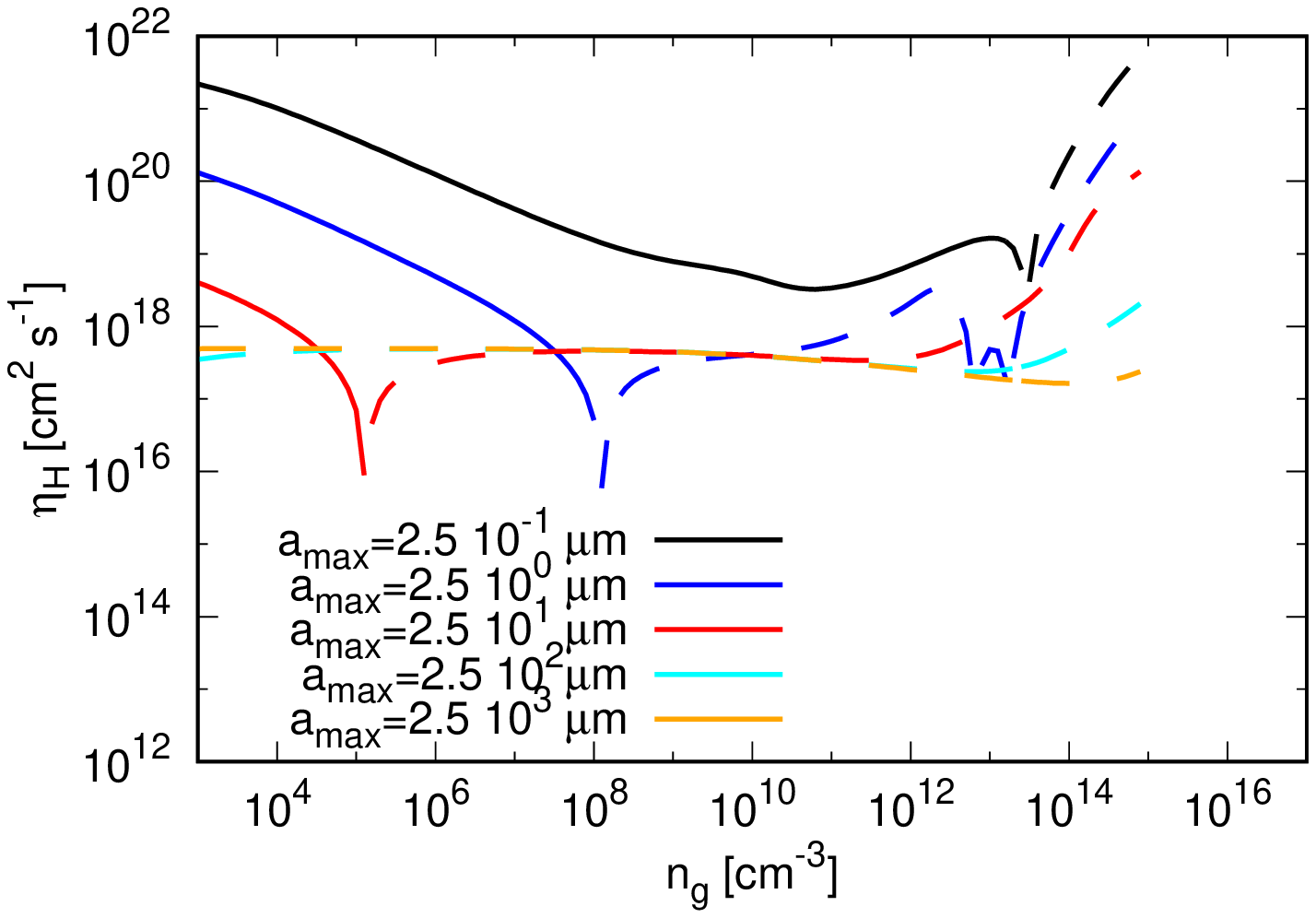}
  \includegraphics[width=40mm,angle=-90]{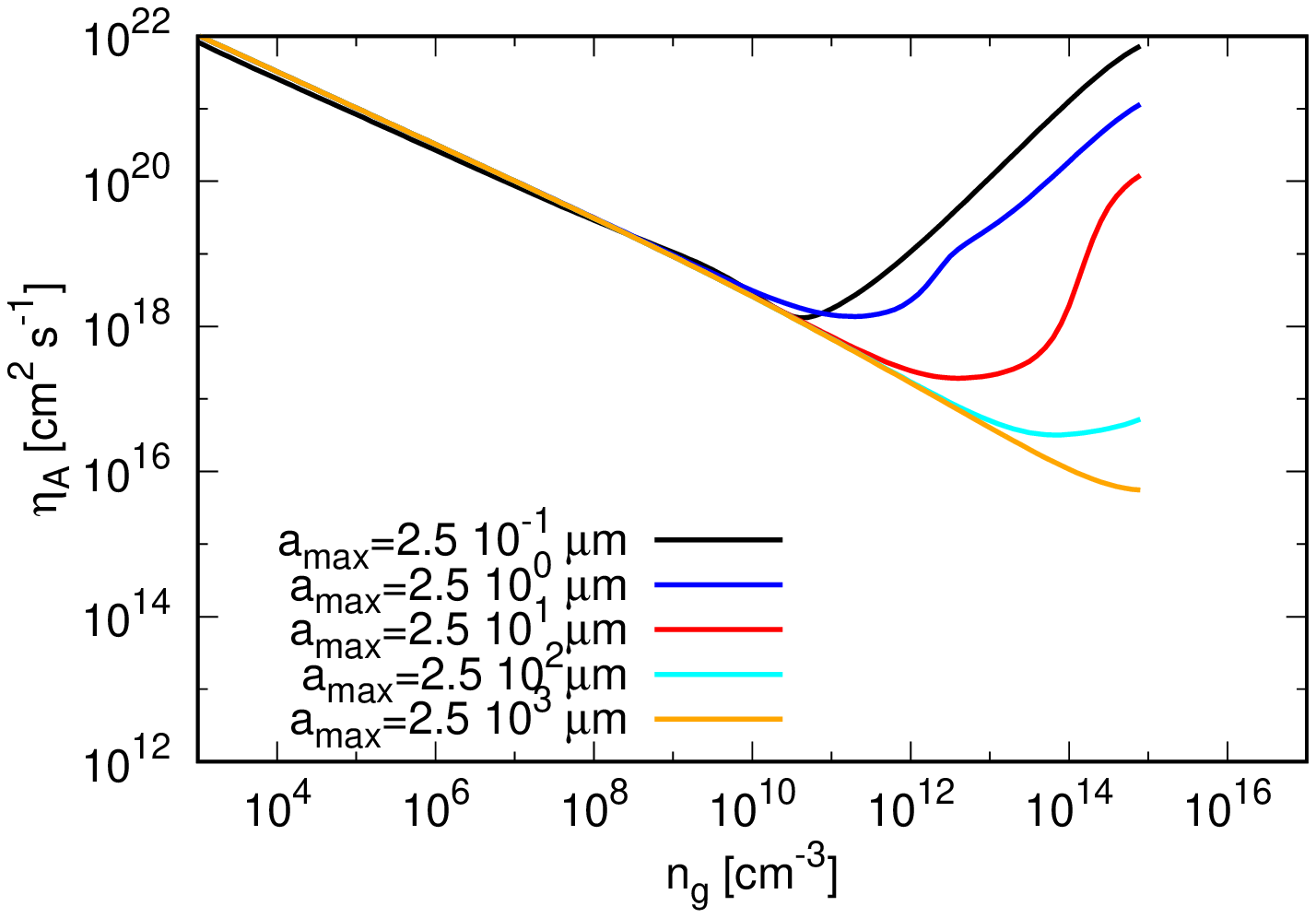}
  \vspace{30mm}
  \caption{
    Same as figure \ref{eta_amin5_q25} but with $a_{\rm min}=100 \nm$ and $q=2.5$.
}
\label{eta_amin100_q25}
\end{figure*}

\begin{figure*}
  \includegraphics[width=40mm,angle=-90]{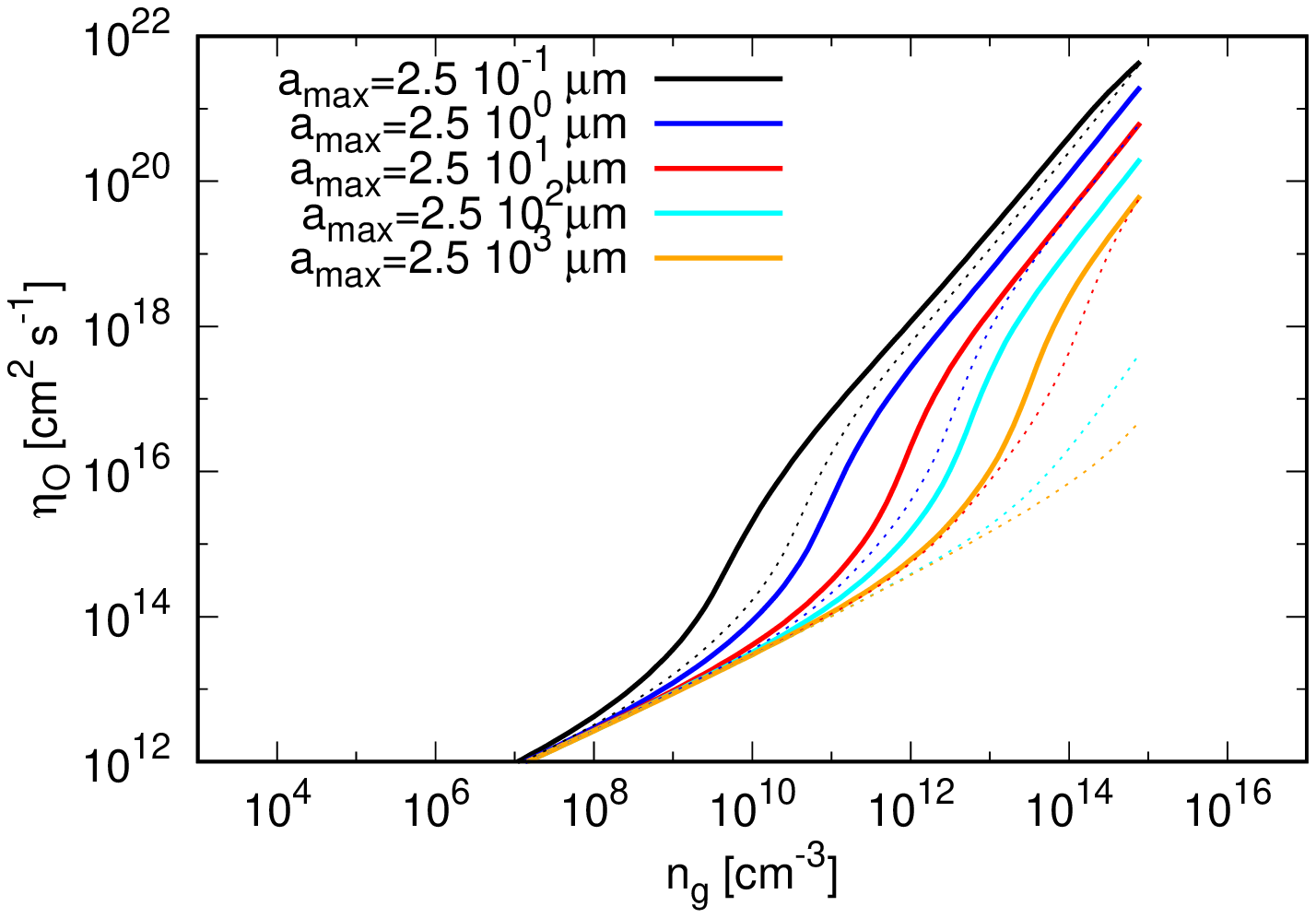}
  \includegraphics[width=40mm,angle=-90]{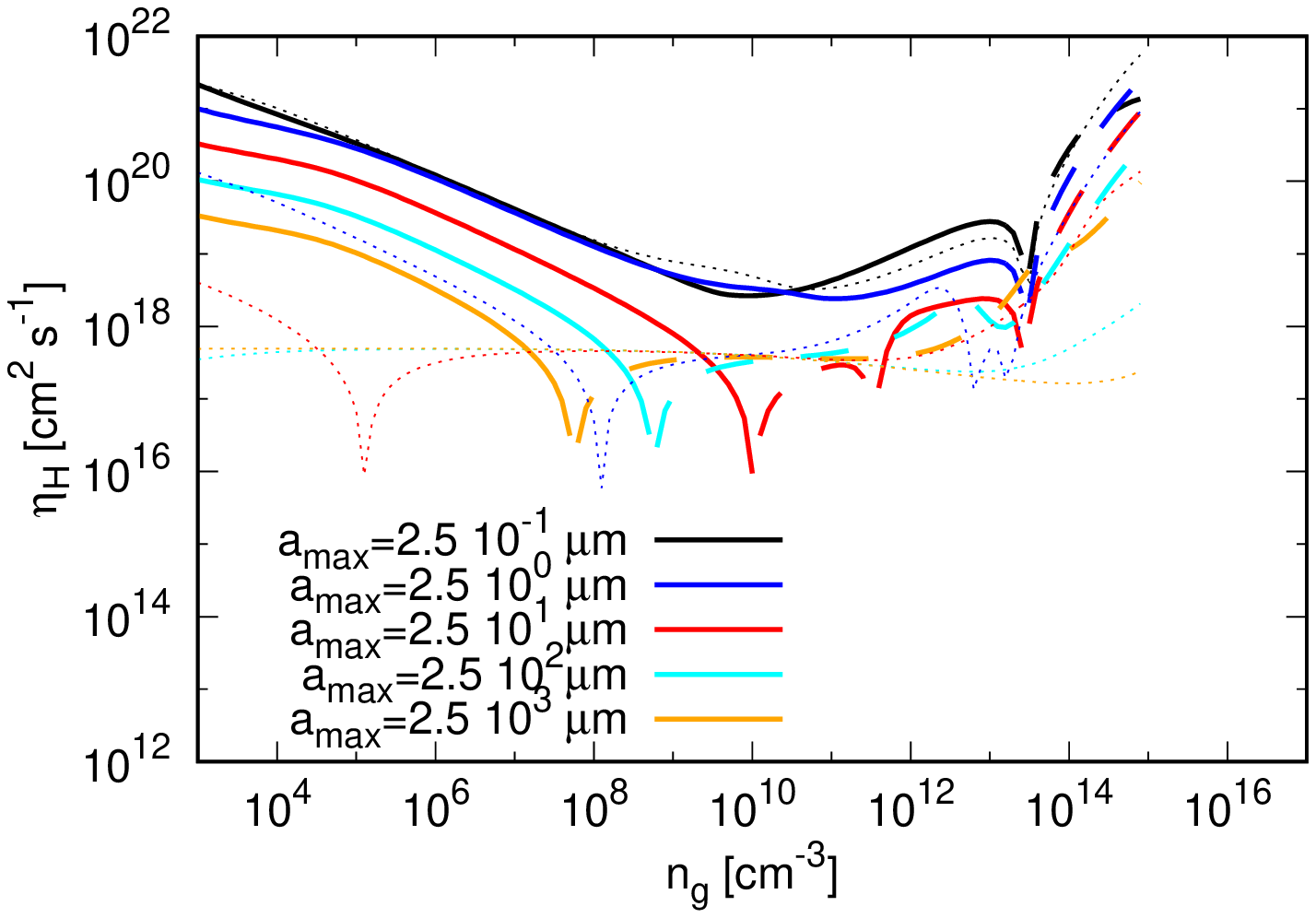}
  \includegraphics[width=40mm,angle=-90]{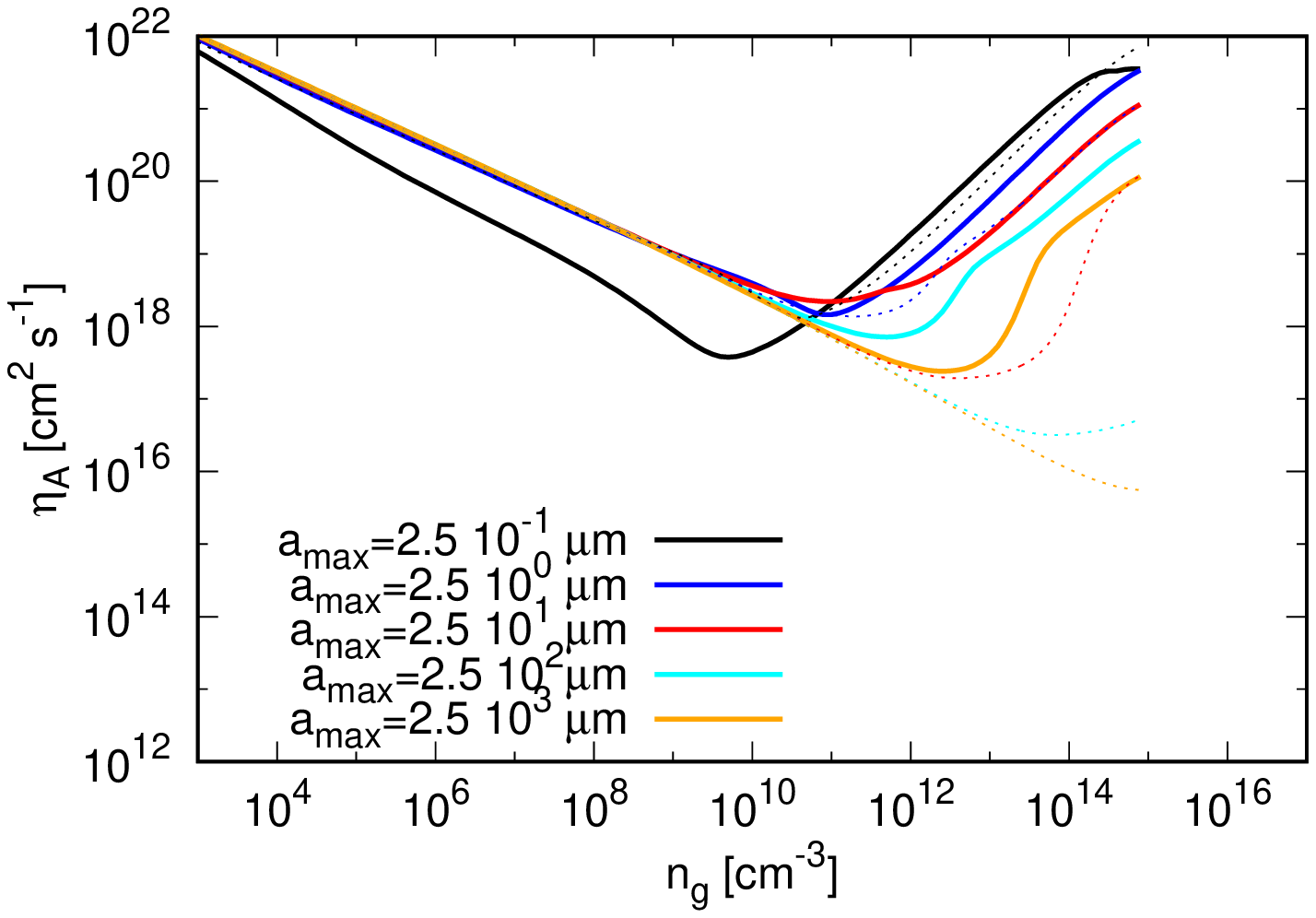}
  \vspace{30mm}
  \caption{
    Same as figure \ref{eta_amin100_q25} but with $a_{\rm min}=100 \nm$ and $q=3.5$.
    Dotted lines show the results with $q=2.5$ for comparison.
}
\label{eta_amin100_q35}
\end{figure*}

Figure \ref{eta_amin5_q25} shows the resistivities with $a_{\rm min}=5 \nm$ and $q=2.5$ for different $a_{\rm max}$.
The figure shows that $\eta_{\rm O}$ tends to decrease with increasing $a_{\rm max}$ aside from $a_{\rm max}=0.25\mum$ to $a_{\rm max}=2.5\mum$ in high density region.
The increase of $\eta_{\rm O}$ and $\eta_{\rm A}$ from $a_{\rm max}=0.25\mum$ to $a_{\rm max}=2.5\mum$ in the high-density region is due to the decrease of dust conductivity.
Then, it can be seen that resistivities converge to  the power law of the form of $\eta_{\rm O, H, A} \propto n_{\rm g}^{p}$ with respective constant power exponent ${\rm p}$ as the dust size increases to $a_{\rm max}=2.5\times 10^3\mum$.
They are power laws determined by the chemical equilibrium of cosmic-ray ionization and gas-phase recombination.
$\eta_{\rm H}$ becomes positive (shown with dashed lines) in $a_{\rm max}> 250 \mum$ almost entire density region.
This is because the relative velocity between ions and electrons determines the Hall current.

Figure \ref{eta_amin5_q35} shows the resistivities with $a_{\rm min}=5 \nm$ and $q=3.5$ for different $a_{\rm max}$.
The figure shows that $\eta_{\rm O}$ tends to decrease with increasing $a_{\rm max}$ in the low-density region, which is due to the decrease of the total cross-section.
On the other hand,  $\eta_{\rm O}$ increases in the high-density region, which is due to the decrease of dust conductivity.
$\eta_{\rm H}$ and $\eta_{\rm A}$  also increase in the high-density region, which is also due to the decrease of dust conductivity there.
The difference between $q=2.5$ (dotted lines) and $q=3.5$ (solid lines) is striking. If we compare $\eta_{\rm O}$ and
$\eta_{\rm A}$ at $10^{15}\ccm$ between figure \ref{eta_amin5_q25}  and \ref{eta_amin5_q35} ,
the difference is more than $10^3$ times greater, which may significantly affect the disk evolution.

Figure \ref{eta_amin100_q25} shows the resistivities with $a_{\rm min}=0.1 \mum$ and $q=2.5$ for different $a_{\rm max}$.
By removing small dust grains, resistivity behavior is simplified because the dust grains themselves are no longer responsible for conductivity
and serve only as adsorber of ions and electrons.  
The figure shows that $\eta_{\rm O}$  tends to decrease with increasing $a_{\rm max}$ in the entire density region.
It can be seen that resistivities converge to the single power law as the dust size increases, which is the same as the figure \ref{eta_amin5_q25}.
Again $\eta_{\rm H}$ becomes positive in $a_{\rm max}> 250 \mum$ almost entire density region, which is also consistent with the figure \ref{eta_amin5_q25}.

Figure \ref{eta_amin100_q35} shows the resistivities with $a_{\rm min}=0.1 \mum$ and $q=3.5$ for different $a_{\rm max}$.
Similar to the result of figure \ref{eta_amin100_q25}, $\eta_{\rm O}$ tends to decrease with increasing $a_{\rm max}$.
However, the decrease is less pronounced because of the contribution of the small dust to the cross-section.
$\eta_{\rm A}$ increases in the low-density region of $n_{\rm g} \lesssim 10^{10} \ccm$ and converges to the single power law as the dust size increases.
On the other hand, it decreases with increasing $a_{\rm max}$ in the high density region of $n_{\rm g} \gtrsim 10^{10} \ccm$. However, again the decrease is less pronounced.
$\eta_{\rm H}$ tends to decrease with increasing  $a_{\rm max}$.

\section{Discussion}
\label{discussion}

In this paper, we investigate the impact of dust size distribution with large dust grains on the magnetic resistivities using the analytic method based
on \citet{1987ApJ...320..803D}. Our test results show that the analytic model can correctly calculate the ionization state from small
$\tau$ (low temperature or small dust) to large $\tau$ (high temperature or large dust).
Therefore, the method is applicable to a dust size distribution that simultaneously contains
small dust with $\tau<1$ and large dust with $\tau>1$, and can be used over a broader class of dust size distribution
than previous study which uses the Gaussian charge distribution such as \citet{2009ApJ...698.1122O,2021ApJ...913..148T}.

The calculation results with large dust grains show that the resistivity tends to decrease with dust growth.
This is particularly true when the dust size power exponent $q$ is $q=2.5$
(i.e., in the case the coagulation process dominates in the dust size evolution,
and only large dust grains are responsible for the dust cross-section).
On the other hand, the decrease is less pronounced when the dust size power exponent $q$ is $q=3.5$,
(i.e., in the case the disruption process dominates in the dust size evolution, and the small dust grains are also responsible for the dust cross-section).
Our results suggest that detailed dust coagulation and fragmentation processes play a crucial role to investigate the impact of non-ideal effects, in particular in the high density region of $10^{10} \ccm$.

Recently, \citet{2021A&A...649A..50M} proposed a similar method that also can be used to calculate magnetic resistivity analytically.
Our numerical tests showed that our method seems to be more robust and applicable over a wide parameter range (e.g., when the dust grains are highly depleted).
When we implemented and tested their algorithm, we found that it did not converge in the limit where the total dust charge goes zero.
This is because their method uses $\psi$ as the basic variable and use equation (A.3) and (A.4) of \citet{2021A&A...649A..50M} to determine $\epsilon$ and
$n_{\rm i}$. However their equation (A.4) becomes singular when $\epsilon=1$ and $\langle \bar{Z} \rangle=0$ i.e.,
gas phase recombination determines the ionization state. That would be a reason why the solution does not converge.

Our model does not include charge neutralization due to grain-grain collisions
which many previous studies have included \citep[e.g.,][]{1990MNRAS.243..103U,2015MNRAS.452..278T,2016A&A...592A..18M}.
One might find this to be a flaw in our model.
However, we would argue here that inclusion of charge neutralization by grain-grain collisions
is debatable and may not necessarily describe a realistic dust charge state in dense region.
This is because, when (sub-)micron-sized small dust particles collide,
the dust particles tends to coalesce and grow rather than bounce
because the collisional velocity of small dust grains tends to be much smaller than their bouncing velocity
\citep{1997ApJ...480..647D,2000PhRvL..85.2426B,2012Icar..218..688W,2015ApJ...798...34G}.

More quantitatively, bouncing threshold velocity for dust grains composed of SiO$_2$
bellow which the collsion results in 50 \% sticking is given as 
\begin{align}
  \Delta v_{\rm stick}=\left(\frac{m_{\rm d}}{m_{\rm th}}\right)^{-5/18} \sim 3.9 \times 10^2 \left(\frac{a_{\rm d}}{10 \nm}\right)^{-15/18} \cms,
\end{align}
where $m_{\rm th}=1.1 \times 10^{-15} ~{\rm g}$ \citep{2012Icar..218..688W}.
On the other hand, the relative velocity of the dust (assuming Brownian motion) is given as
\begin{align}
  \Delta v_{\rm Brown}=\sqrt{\frac{16 k_{\rm B} T}{\pi m_{\rm d}}} \sim 91 \left(\frac{a_d}{10 \nm}\right)^{-3/2} \left(\frac{T}{100 {\rm K}}\right)^{1/2} \cms
\end{align}
By solving the inequality of $\Delta v_{\rm stick}>\Delta v_{\rm Brown}$, we can conclude that the silicate dust grain with $a_{\rm d}>1.2  (T/100 {\rm K})^{3/4} \nm$
tends to stick rather than bounce, which is much smaller than the minimum size of MRN size distribution ($a_{\rm min}=5 \nm$).
Note also icy or porous dust grains are even more sticky \citep{2009ApJ...702.1490W}.
Hence, grain-grain collisions lead to dust growth and a change in the dust size distribution rather than bounce. 

Therefore, the approximation that ignores charge neutralization due to grain-grain collisions
is not necessarily a flaw in our model, but rather a difference
in the approximation of how we view the dust collision process.
Note also that neglecting grain-grain neutralization does not cause the change on the resistivities
when we consider the large dust (e.g., $\gtrsim 1\mum$),
which is the main subject of current and our subsequent studies.
This is because the charged dust grains determines the resistivities only
when there are sufficient sub-micron dust grains ($\lesssim 100\nm$).

  Several studies  have investigated the effect of dust size distribution on magnetic resistivities.
  \citep[e.g.,][]{2018MNRAS.478.2723Z,2020A&A...643A..17G}. Our results seems to be consistent with these studies.
  For example,  in comparison with \citet{2018MNRAS.478.2723Z}, figure \ref{eta_amin5_q35} shows that
  $\eta_{\rm O}$ increases 
  from $a_{\rm max}=2.5 \times 10^{-1} \mum$ to $a_{\rm max}=2.5 \mum$ at $n_g \sim 10^{14} \ccm$, and decreases
  at  $n_g \sim 10^{12} \ccm$.
  $\eta_{\rm A}$ increases in $n_g<10^{10} \ccm$ and converged
  to the power law of $n_g^{-1/2}$ \citep{1983ApJ...273..202S} as dust size increases.
  $\eta_{\rm H}$ becomes positive and almost constant as dust size increases.
  Although \citet{2018MNRAS.478.2723Z} treats chemical reactions in more detail
  and the dust size distribution considered is different
  (they changed $a_{\rm min}$ instead of $a_{\rm max}$),
  these trends are largely consistent with their results (see their figure 5).

  On the other hand, the comparison with \citet{2020A&A...643A..17G}
  is difficult because their dust size distribution changes as density increases
  and they also include non-thermal dust drift due to ambipolar diffusion.
  However, the following consistent trends can be observed.
  Our figure \ref{eta_amin100_q25} shows that smaller power exponent $q$
  (meaning that small dust aggregates are less abundant) causes significant decreases of $\eta_{\rm O}$ and increase of $\eta_{\rm A}$ around intermediate density ($n_g \sim 10^9 \ccm$).
  On the other hand, figure 9 of \citet{2020A&A...643A..17G}
  shows that larger $V_{\rm AD}$ (causing the removal of small dust) results in 
  the decrease of $\eta_{\rm O}$ and the increase of $\eta_{\rm A}$ around the intermediate density.
  These points are consistent.

The method described in this paper is less computationally expensive  than conventional chemical reaction network calculations and easily converges to the solutions
because it only requires to perform one-dimensional Newton-Raphson method twice (in determining $\psi$ and $\epsilon$).  Our method typically requires only a few (typically 1-4 times) iterations for each Newton-Raphson calculation to obtain a result.
Therefore, it can be easily used in 3D simulations with negligible computational costs.
Upon request, we will provide a sample implementation of our method to the readers.

We plan to use the analytic model in this paper in 3D MHD simulations of disk formation and evolution which incorporates dust growth.

\section*{Acknowledgments}

This work is supported by JSPS KAKENHI  grant number 18H05437, 18K13581, 18K03703.

\appendix

\section{Comparison with \citet{2021A&A...649A..50M}}

\begin{figure*}
  \includegraphics[width=50mm,angle=-90]{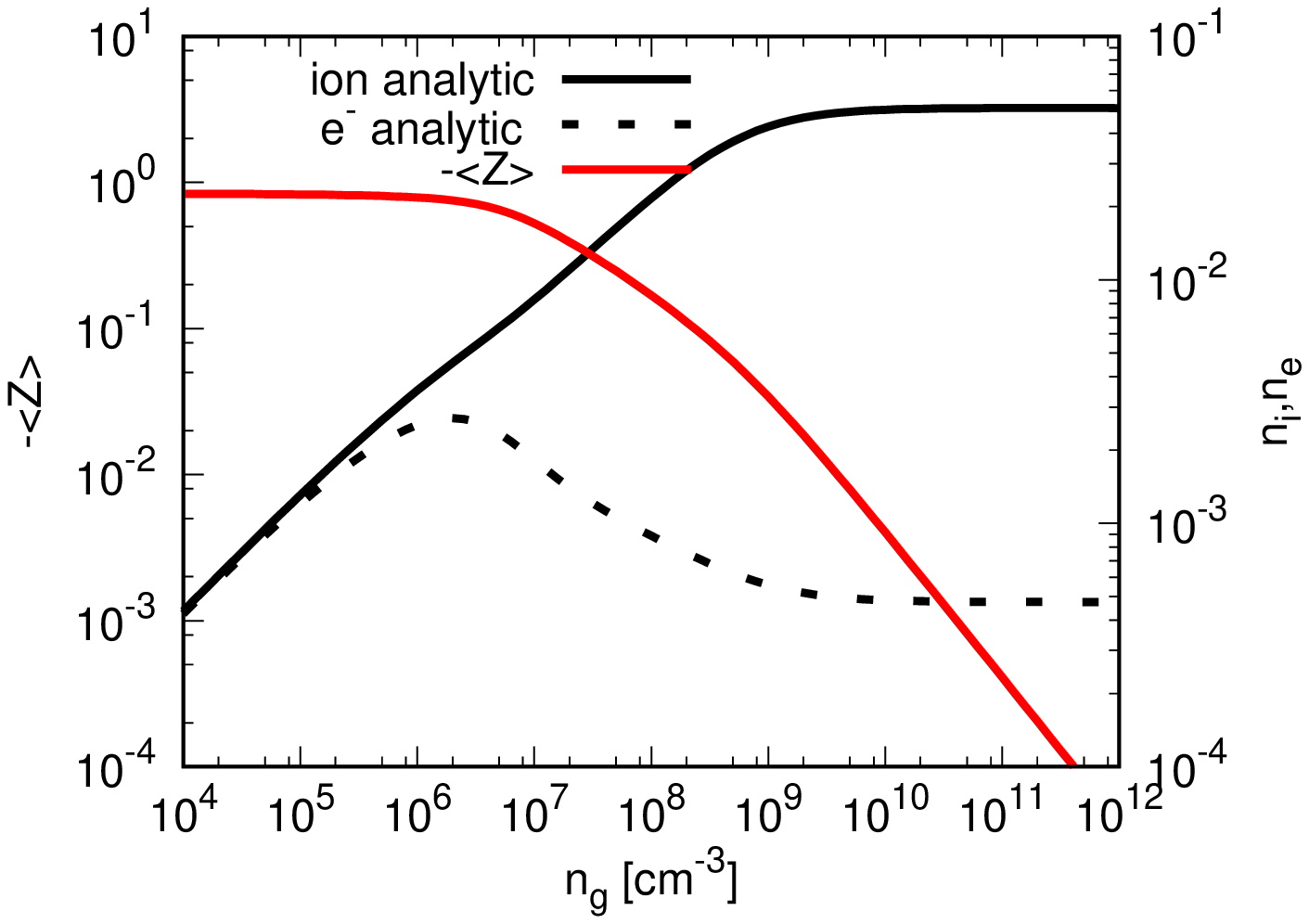}
  \vspace{40mm}
  \caption{
    The number density of  ion (black solid), electron  (black dashed),
    and mean dust charge (thick red) from the analytic calculation.
    The dust size distribution is assumed to be MRN size distribution.
    In this plot, we assume the parameters of figure 2 of \citet{2021A&A...649A..50M}.
}
\label{abundance_marchand}
\end{figure*}

For comparison with the previous study by \citet{2021A&A...649A..50M},
  we plot the number density of ions and electrons and $-\langle Z\rangle$ calculated from our analytic model
  in figure \ref{abundance_marchand}.
  In this figure, we assume the parameters of \citet{2021A&A...649A..50M}.
  This figure can be directly compared to figure 2 of  \citet{2021A&A...649A..50M}.
  The results in this figure are in good agreement with their results.
  Quantitatively, we confirmed that  $n_{\rm i}/n_{\rm e} $ converges
  to $\Theta \equiv s_{\rm e} (m_{\rm i}/m_{\rm e})^{1/2}=107$ in $n_{\rm g} \gtrsim 10^{12} \ccm$,
  which is also in good agreement with their results.

\bibliography{article}

\begin{thebibliography}{}
\expandafter\ifx\csname natexlab\endcsname\relax\def\natexlab#1{#1}\fi

\bibitem[{{Blum} {et~al.}(2000){Blum}, {Wurm}, {Kempf}, {Poppe}, {Klahr},
  {Kozasa}, {Rott}, {Henning}, {Dorschner}, {Schr{\"a}pler}, {Keller},
  {Markiewicz}, {Mann}, {Gustafson}, {Giovane}, {Neuhaus}, {Fechtig},
  {Gr{\"u}n}, {Feuerbacher}, {Kochan}, {Ratke}, {El Goresy}, {Morfill},
  {Weidenschilling}, {Schwehm}, {Metzler}, \& {Ip}}]{2000PhRvL..85.2426B}
{Blum}, J., {Wurm}, G., {Kempf}, S., {et~al.} 2000, \prl, 85, 2426

\bibitem[{{Dominik} \& {Tielens}(1997)}]{1997ApJ...480..647D}
{Dominik}, C., \& {Tielens}, A.~G.~G.~M. 1997, \apj, 480, 647

\bibitem[{{Draine} \& {Sutin}(1987)}]{1987ApJ...320..803D}
{Draine}, B.~T., \& {Sutin}, B. 1987, \apj, 320, 803

\bibitem[{{Guillet} {et~al.}(2020){Guillet}, {Hennebelle}, {Pineau des
  For{\^e}ts}, {Marcowith}, {Commer{\c{c}}on}, \&
  {Marchand}}]{2020A&A...643A..17G}
{Guillet}, V., {Hennebelle}, P., {Pineau des For{\^e}ts}, G., {et~al.} 2020,
  \aap, 643, A17

\bibitem[{{Gundlach} \& {Blum}(2015)}]{2015ApJ...798...34G}
{Gundlach}, B., \& {Blum}, J. 2015, \apj, 798, 34

\bibitem[{Hindmarsh {et~al.}(2005)Hindmarsh, Brown, Grant, Lee, Serban,
  Shumaker, \& Woodward}]{hindmarsh2005sundials}
Hindmarsh, A.~C., Brown, P.~N., Grant, K.~E., {et~al.} 2005, ACM Transactions
  on Mathematical Software (TOMS), 31, 363

\bibitem[{{Koga} {et~al.}(2019){Koga}, {Tsukamoto}, {Okuzumi}, \&
  {Machida}}]{2019MNRAS.484.2119K}
{Koga}, S., {Tsukamoto}, Y., {Okuzumi}, S., \& {Machida}, M.~N. 2019, \mnras,
  484, 2119

\bibitem[{{Kuffmeier} {et~al.}(2017){Kuffmeier}, {Haugb{\o}lle}, \&
  {Nordlund}}]{2017ApJ...846....7K}
{Kuffmeier}, M., {Haugb{\o}lle}, T., \& {Nordlund}, {\r{A}}. 2017, \apj, 846, 7

\bibitem[{{Machida} \& {Basu}(2019)}]{2019ApJ...876..149M}
{Machida}, M.~N., \& {Basu}, S. 2019, \apj, 876, 149

\bibitem[{{Machida} {et~al.}(2007){Machida}, {Inutsuka}, \&
  {Matsumoto}}]{2007ApJ...670.1198M}
{Machida}, M.~N., {Inutsuka}, S., \& {Matsumoto}, T. 2007, \apj, 670, 1198

\bibitem[{{Machida} {et~al.}(2011){Machida}, {Inutsuka}, \&
  {Matsumoto}}]{2011PASJ...63..555M}
---. 2011, \pasj, 63, 555

\bibitem[{{Machida} {et~al.}(2008){Machida}, {Inutsuka}, \&
  {Matsumoto}}]{2008ApJ...676.1088M}
{Machida}, M.~N., {Inutsuka}, S.-i., \& {Matsumoto}, T. 2008, \apj, 676, 1088

\bibitem[{{Marchand} {et~al.}(2021){Marchand}, {Guillet}, {Lebreuilly}, \& {Mac
  Low}}]{2021A&A...649A..50M}
{Marchand}, P., {Guillet}, V., {Lebreuilly}, U., \& {Mac Low}, M.~M. 2021,
  \aap, 649, A50

\bibitem[{{Marchand} {et~al.}(2016){Marchand}, {Masson}, {Chabrier},
  {Hennebelle}, {Commer{\c c}on}, \& {Vaytet}}]{2016A&A...592A..18M}
{Marchand}, P., {Masson}, J., {Chabrier}, G., {et~al.} 2016, \aap, 592, A18

\bibitem[{{Marchand} {et~al.}(2020){Marchand}, {Tomida}, {Tanaka},
  {Commer{\c{c}}on}, \& {Chabrier}}]{2020ApJ...900..180M}
{Marchand}, P., {Tomida}, K., {Tanaka}, K. E.~I., {Commer{\c{c}}on}, B., \&
  {Chabrier}, G. 2020, \apj, 900, 180

\bibitem[{{Masson} {et~al.}(2016){Masson}, {Chabrier}, {Hennebelle}, {Vaytet},
  \& {Commer{\c c}on}}]{2016A&A...587A..32M}
{Masson}, J., {Chabrier}, G., {Hennebelle}, P., {Vaytet}, N., \& {Commer{\c
  c}on}, B. 2016, \aap, 587, A32

\bibitem[{{Mathis} {et~al.}(1977){Mathis}, {Rumpl}, \&
  {Nordsieck}}]{1977ApJ...217..425M}
{Mathis}, J.~S., {Rumpl}, W., \& {Nordsieck}, K.~H. 1977, \apj, 217, 425

\bibitem[{{McElroy} {et~al.}(2013){McElroy}, {Walsh}, {Markwick}, {Cordiner},
  {Smith}, \& {Millar}}]{2013A&A...550A..36M}
{McElroy}, D., {Walsh}, C., {Markwick}, A.~J., {et~al.} 2013, \aap, 550, A36

\bibitem[{{Miyake} \& {Nakagawa}(1993)}]{1993Icar..106...20M}
{Miyake}, K., \& {Nakagawa}, Y. 1993, \icarus, 106, 20

\bibitem[{{Nakano} {et~al.}(2002){Nakano}, {Nishi}, \&
  {Umebayashi}}]{2002ApJ...573..199N}
{Nakano}, T., {Nishi}, R., \& {Umebayashi}, T. 2002, \apj, 573, 199

\bibitem[{{Okuzumi}(2009)}]{2009ApJ...698.1122O}
{Okuzumi}, S. 2009, \apj, 698, 1122

\bibitem[{{Ormel} \& {Cuzzi}(2007)}]{2007A&A...466..413O}
{Ormel}, C.~W., \& {Cuzzi}, J.~N. 2007, \aap, 466, 413

\bibitem[{{Pinto} \& {Galli}(2008)}]{2008A&A...484...17P}
{Pinto}, C., \& {Galli}, D. 2008, \aap, 484, 17

\bibitem[{{Shu}(1983)}]{1983ApJ...273..202S}
{Shu}, F.~H. 1983, \apj, 273, 202

\bibitem[{{Tomida} {et~al.}(2017){Tomida}, {Machida}, {Hosokawa}, {Sakurai}, \&
  {Lin}}]{2017ApJ...835L..11T}
{Tomida}, K., {Machida}, M.~N., {Hosokawa}, T., {Sakurai}, Y., \& {Lin}, C.~H.
  2017, \apjl, 835, L11

\bibitem[{{Tomida} {et~al.}(2015){Tomida}, {Okuzumi}, \&
  {Machida}}]{2015ApJ...801..117T}
{Tomida}, K., {Okuzumi}, S., \& {Machida}, M.~N. 2015, \apj, 801, 117

\bibitem[{{Tomida} {et~al.}(2013){Tomida}, {Tomisaka}, {Matsumoto}, {Hori},
  {Okuzumi}, {Machida}, \& {Saigo}}]{2013ApJ...763....6T}
{Tomida}, K., {Tomisaka}, K., {Matsumoto}, T., {et~al.} 2013, \apj, 763, 6

\bibitem[{{Tsukamoto} {et~al.}(2015{\natexlab{a}}){Tsukamoto}, {Iwasaki},
  {Okuzumi}, {Machida}, \& {Inutsuka}}]{2015ApJ...810L..26T}
{Tsukamoto}, Y., {Iwasaki}, K., {Okuzumi}, S., {Machida}, M.~N., \& {Inutsuka},
  S. 2015{\natexlab{a}}, \apjl, 810, L26

\bibitem[{{Tsukamoto} {et~al.}(2015{\natexlab{b}}){Tsukamoto}, {Iwasaki},
  {Okuzumi}, {Machida}, \& {Inutsuka}}]{2015MNRAS.452..278T}
---. 2015{\natexlab{b}}, \mnras, 452, 278

\bibitem[{{Tsukamoto} {et~al.}(2021{\natexlab{a}}){Tsukamoto}, {Machida}, \&
  {Inutsuka}}]{2021ApJ...913..148T}
{Tsukamoto}, Y., {Machida}, M.~N., \& {Inutsuka}, S. 2021{\natexlab{a}}, \apj,
  913, 148

\bibitem[{{Tsukamoto} {et~al.}(2021{\natexlab{b}}){Tsukamoto}, {Machida}, \&
  {Inutsuka}}]{2021ApJ...920L..35T}
{Tsukamoto}, Y., {Machida}, M.~N., \& {Inutsuka}, S.-i. 2021{\natexlab{b}},
  \apjl, 920, L35

\bibitem[{{Tsukamoto} {et~al.}(2020){Tsukamoto}, {Machida}, {Susa}, {Nomura},
  \& {Inutsuka}}]{2020ApJ...896..158T}
{Tsukamoto}, Y., {Machida}, M.~N., {Susa}, H., {Nomura}, H., \& {Inutsuka}, S.
  2020, \apj, 896, 158

\bibitem[{{Umebayashi} \& {Nakano}(1990)}]{1990MNRAS.243..103U}
{Umebayashi}, T., \& {Nakano}, T. 1990, \mnras, 243, 103

\bibitem[{{Vaytet} {et~al.}(2018){Vaytet}, {Commer{\c{c}}on}, {Masson},
  {Gonz{\'a}lez}, \& {Chabrier}}]{2018A&A...615A...5V}
{Vaytet}, N., {Commer{\c{c}}on}, B., {Masson}, J., {Gonz{\'a}lez}, M., \&
  {Chabrier}, G. 2018, \aap, 615, A5

\bibitem[{{Wada} {et~al.}(2009){Wada}, {Tanaka}, {Suyama}, {Kimura}, \&
  {Yamamoto}}]{2009ApJ...702.1490W}
{Wada}, K., {Tanaka}, H., {Suyama}, T., {Kimura}, H., \& {Yamamoto}, T. 2009,
  \apj, 702, 1490

\bibitem[{{Wardle}(2007)}]{2007Ap&SS.311...35W}
{Wardle}, M. 2007, \apss, 311, 35

\bibitem[{{Weidling} {et~al.}(2012){Weidling}, {G{\"u}ttler}, \&
  {Blum}}]{2012Icar..218..688W}
{Weidling}, R., {G{\"u}ttler}, C., \& {Blum}, J. 2012, \icarus, 218, 688

\bibitem[{{Wurster} {et~al.}(2016){Wurster}, {Price}, \&
  {Bate}}]{2016MNRAS.457.1037W}
{Wurster}, J., {Price}, D.~J., \& {Bate}, M.~R. 2016, \mnras, 457, 1037

\bibitem[{{Xu} \& {Kunz}(2021)}]{2021MNRAS.502.4911X}
{Xu}, W., \& {Kunz}, M.~W. 2021, \mnras, 502, 4911

\bibitem[{{Zhao} {et~al.}(2018{\natexlab{a}}){Zhao}, {Caselli}, \&
  {Li}}]{2018MNRAS.478.2723Z}
{Zhao}, B., {Caselli}, P., \& {Li}, Z.-Y. 2018{\natexlab{a}}, \mnras, 478, 2723

\bibitem[{{Zhao} {et~al.}(2018{\natexlab{b}}){Zhao}, {Caselli}, {Li}, \&
  {Krasnopolsky}}]{2018MNRAS.473.4868Z}
{Zhao}, B., {Caselli}, P., {Li}, Z.-Y., \& {Krasnopolsky}, R.
  2018{\natexlab{b}}, \mnras, 473, 4868

\bibitem[{{Zhao} {et~al.}(2016){Zhao}, {Caselli}, {Li}, {Krasnopolsky},
  {Shang}, \& {Nakamura}}]{2016MNRAS.460.2050Z}
{Zhao}, B., {Caselli}, P., {Li}, Z.-Y., {et~al.} 2016, \mnras, 460, 2050

\end{thebibliography}

\end{document}